\documentclass[preprint,aps,showpacs]{revtex4}
\usepackage{graphicx}
\usepackage{amsmath,mathrsfs,epsfig}
\begin{document}
\bibliographystyle{revtex}
\title{Exact soliton solutions of coupled
nonlinear Schr\"odinger equations: Shape changing collisions, logic gates and
partially coherent solitons}

\author{T. Kanna and M. Lakshmanan}
\email{cnld123@eth.net}

\affiliation{Centre for Nonlinear Dynamics, Department of Physics,
Bharathidasan University, Tiruchirapalli 620 024, India}

\vspace{2pt}
\newcommand{\be}{\begin{equation}}
\newcommand{\ee}{\end{equation}}
\newcommand{\bear}{\begin{eqnarray}}
\newcommand{\eear}{\end{eqnarray}}
\newcommand{\bes}{\begin{subequations}}
\newcommand{\ees}{\end{subequations}}
\newcommand{\al}{\alpha}
\newcommand{\ba}{\beta}
\newcommand{\ga}{\gamma}
\newcommand{\del}{\delta}
\newcommand{\sech}{\mbox{ sech}}

\begin{abstract}
The novel dynamical features underlying soliton interactions in coupled
nonlinear Schr{\"o}dinger equations, which model multimode wave propagation
under varied physical situations in nonlinear optics, are studied. In this
paper, by explicitly constructing multisoliton solutions (upto four-soliton
solutions) for two coupled  and
arbitrary $N$-coupled nonlinear Schr{\"o}dinger equations using the Hirota
bilinearization method, we bring out clearly the various features underlying the
fascinating shape changing (intensity redistribution) collisions of solitons, including changes
in amplitudes, phases and relative separation distances, and the very many
possibilities of energy redistributions among the modes of solitons. However in
this multisoliton collision process the pair-wise collision nature is shown to be preserved in spite of
the changes in the amplitudes and phases of the solitons. Detailed
asymptotic analysis also shows that when solitons undergo multiple collisions,
there exists the exciting possibility of shape restoration of atleast one
soliton during interactions of more than two solitons represented by three 
and higher order soliton solutions. From application point of view, we have
shown from the asymptotic expressions how the amplitude (intensity)
redistribution can be written as a generalized linear fractional transformation
for the $N$-component case. Also we indicate how the multisolitons can be
reinterpreted as various logic gates for suitable choices of the soliton
parameters, leading to possible multistate logic. 
In addition, we point out that the various
recently studied partially coherent solitons are just special cases of the bright soliton solutions
exhibiting shape changing collisions, thereby explaining their variable profile
and shape variation in collision process.
\end{abstract}

\maketitle

\section{Introduction}
The study of coupled nonlinear Schr{\"o}dinger (CNLS) equations is receiving a 
great deal of attention in recent years due to their appearance as modelling 
equations in diverse areas of physics like nonlinear optics [1],  
including optical communications [2], bio-physics [3], multicomponent 
Bose-Einstein condensates at 
zero temperature [4], etc.  To be specific, soliton type pulse propagation in
multimode fibers [1] and in fiber arrays [5] is governed by a set of $N$-CNLS equations
which is not integrable in general. However, it becomes integrable for specific
choice of parameters [6,7]. On the other hand, the recent studies on the
coherent [8] and incoherent [9] beam propagation in photorefractive media,
which can exhibit high nonlinearity with extremely low optical power,
necessitate intense study of  CNLS equations both integrable and nonintegrable.
 The first experimental
observation of the so-called partially incoherent solitons with the excitation
of a light bulb in a photorefractive medium [10] has made this study even more interesting. In
this context of beam propagation in a Kerr like photorefractive medium, the
governing equations are a set of $N$-CNLS equations [11,12].

We consider the following $N$-CNLS equations of Manakov type [13] for our study,
\bear
iq_{jz}+q_{jtt}+2\mu\sum_{p=1}^N|q_p|^2q_j=0,\;\;j=1,2,...,N,
\eear 
where $q_j$ is the envelope in the $j$th mode, $z$ and $t$ represent the 
normalized distance along the fiber and the retarded time respectively, in the 
context of soliton propagation in multimode fibers. In the case of fiber 
arrays $q_j$ corresponds to the $j$th core. 
Here $2\mu$ gives the strength of the nonlinearity. In the framework of 
N self trapped mutually incoherent wavepackets propagation in Kerr-like photorefractive 
media [11,12], $q_j$ is the $j$th component of the beam, $z$ and $t$ represent 
the coordinates along the direction of propagation and the transverse 
coordinate, respectively. The interesting property of the
 $N$-CNLS equations of the form (1) is that they are integrable equations and 
possess soliton solutions.

It is obvious from Eq. (1) that for $N=1$ it corresponds to the standard envelope 
soliton possessing integrable nonlinear Schr{\"o}dinger equation, governing 
intense optical pulse propagation through single mode optical fiber [1,14]. For 
$N=2$ case, it reduces to the celebrated Manakov model [13] describing intense
electromagnetic pulse propagation in birefringent fiber. Manakov himself has
made a detailed asymptotic analysis of the inverse scattering problem associated
with the system (1) for $N=2$ and identified changes in the polarization vector
[13]. However no explicit two soliton expression was given there. Very recently, 
Radhakrishnan, Lakshmanan and Hietarinta have obtained the bright one- and 
two-soliton solutions for this case [15] and have revealed certain novel shape 
changing (intensity redistribution) collision properties. These Manakov solitons
have been observed recently in AlGaAs planar waveguides [16] and precisely this
kind of energy exchanging (shape changing) collisions has been experimentally
demonstrated in [17]. The results of Ref. [15] have led Jakubowski, Squier 
and Steiglitz [18] to express the energy redistributions as linear fractional 
transformations so as to construct logic gates. Later, Steiglitz [19] explicitly 
constructed such logic gates including the universal NAND gate, 
based on the shape changing collision property, and hence pointed out the 
possibility of designing an all optical computer equivalent to a Turing 
machine, at least in a mathematical sense. However, results are scarce for 
$N\geq 2$ case of Eq. (1) though they are of considerable physical importance 
as mentioned earlier. 

The novel shape changing collision property exhibited by the 2-CNLS equations, 
which has not been observed in general in any other simpler $(1+1)$ dimensional 
integrable system, requires a detailed analysis to identify the various 
possibilities and the underlying potential technological applications. In a very
recent letter [20], the present authors have  studied the multicomponent N-CNLS 
equations and shown that shape changing collisions occur here also with more 
possibilities of energy redistribution. It has also been briefly pointed out 
that the much discussed partially coherent solitons (PCSs) [11,12], which are 
of variable shape, namely 2-PCS, 3-PCS,..., $N$-PCS, are special cases of the 2-soliton, 
3-soliton,..., $N$-soliton solutions of the 2-CNLS, 3-CNLS,..., $N$-CNLS equations,
respectively. The understanding of variable shapes [11,12] of these recently 
experimentally observed partially coherent solitons [21] in 
photorefractive medium and their interesting collision behaviour will be 
fascilitated by obtaining the higher order soliton solutions of the 2-CNLS and 
the $N$-CNLS (N$\geq 2$) equations. 

In this paper, we wish to undertake a detailed analysis of the 
dynamical features associated with soliton interactions in multicomponent $N$-CNLS
equations. There exists numerous interesting phenomena which one has to pay attention in
order to realize the full potentialities of these equations and the underlying
novel soliton dynamics. Some of the important aspects include the following 
among others:
\begin{enumerate}
\item
Explicit expressions for multisoliton solutions in multicomponent CNLS 
equations useful for analysis of interactions (as against formal expressions).
\item
Novel soliton interactions involving shape changing collisions.
\item
Dependence of shape changes and relative separation distances on amplitudes
of the colliding solitons.
\item
Identification of different possibilities of energy redistributions among the
different modes of the soliton during collision and obtaining generalized linear
fractional transformations. 
\item
State restoring properties in multisoliton solutions.
\item
Existence of partially coherent solitons, stationary and moving, as special
cases of the above multisoliton solutions.
\item
Identification of multisoliton solutions as logic gates in multicomponent CNLS
equations.
\end{enumerate}

The present paper will be essentially devoted to the understanding of
multisoliton interactions in N-CNLS equations, and its application in
constructing logic gates and in identifying partially coherent solitons as
special cases of multisoliton solutions. In particular, in the present paper, we will deduce explicit
expressions for multisoliton solutions (upto four-soliton solution), which can
be easily generalized to arbitrary soliton case, 2-CNLS and then for arbitrary
$N$-CNLS equations. To start with, we will briefly consider the two-soliton
solution to bring out the shape changing nature of soliton collisions, which can
be quantified in terms of generalized linear fractional transformations (LFTs),
and identify the changes in amplitudes, phases and relative
separation distances among the solitons by carrying out appropriate asymptotic
analysis.  However the standard (shape preserving) elastic collisions can occur 
for specific choice of soliton parameters (initial conditions). More
interestingly, we also point out that when more than two solitons collide 
successively, say three solitons, there exists the exciting possibility of
restoration of the shape of one of the three solitons leaving the other two
undergoing shape changes and we prove that the underlying soliton interaction is
pair-wise. We give explicit conditions for the shape restoring
property. Extending this analysis, one can easily check that in an $M$-soliton
collision, it is possible to restore the states of $(M-2)$ solitons after
collisions. Such possibilities lead to the construction of optical logic gates
of different types and generalized linear fractional transformations, 
as we will show in this manuscript. 

This paper is organized as follows. In Sec. II we briefly present the 
bilinearization procedure for the $N$-CNLS equations. 
Explicit multisoliton solutions (upto four) of the 2-CNLS equations are obtained in 
Sec. III. 
Then the generalization of these multisoliton solutions 
to $N$-CNLS equations is given in Sec. IV. 
The two soliton collision properties of 2-CNLS  and
their generalization to $N$-CNLS equations are 
studied in Sec. V. In Sec. VI, we present a systematic procedure to identify the
intensity redistribution among $N$ modes in terms of a generalized linear
fractional transformation which is the precursor to develop logic gates
without interconnecting discrete components [18]. The interesting feature 
of the higher order soliton solutions, namely the pair-wise nature of collision of
solitons, and
the shape restoration property of the state of one soliton only in a three soliton
collision process are presented in Sec. VII. In Sec. VIII we introduce the
possibility of looking at the bright soliton solutions as logic gates, as an
alternate point of view. Then in Sec. IX we demonstrate explicitly that for specific
choices of the parameters of the
bright soliton solutions various PCSs reported in the literature result. 
The collision properties of PCSs
 and the salient features of multisoliton complexes are also
discussed. Sec. X is allotted for
conclusion. Also in Appendix A we present the explicit form of
the four-soliton solution.
\section{Bilinearization}
The set of $N$-CNLS equations (1) has been  found to be completely integrable 
[6,7] and admits exact  
bright soliton solutions. Their explicit forms can be obtained  by using the 
Hirota's bilinearization method [22], which is straightforward. Any of the other
soliton producing methodologies in principle is equally applicable, however,
this paper is not concerned with the relative merits of the various
methodologies. 

To start with, we make the bilinearizing transformation (which can be identified
systematically from the Laurent expansion [6])
\begin{equation}
q_j=\frac{g^{(j)}}{f},\;\; j=1,2,...,N,
\end{equation}
to Eq. (1). This results in the following set of bilinear equations,
\bes
\bear
(iD_z+D_t^2)g^{(j)}.f&=& 0,\;\; j=1,2,...,N,\\
D_t^2f.f&=& 2\mu\sum_{n=1}^N{g^{(n)}g^{(n)*}}, 
\eear
where $*$ denotes the complex conjugate, $g^{(j)}$'s  are
 complex functions, while
$f(z,t)$ is a real function and the Hirota's bilinear operators 
$D_z$ and $D_t$ are defined by 
\bear
D_z^nD_t^m(a.b)=\left(\frac{\partial}{\partial z}-\frac{\partial}{\partial z'}
\right)^n
\left(\frac{\partial}{\partial t}-\frac{\partial}{\partial t'}
\right)^ma(z,t)b(z',t')
\Big{\vert}_{(z=z', t=t')}.
\eear
\ees
The above set of equations can be solved by introducing the following 
power series expansions for $g^{(j)}$'s and $f$:
\begin{subequations}
\begin{eqnarray}
g^{(j)} & = & \chi g_1^{(j)} + \chi^3 g_3^{(j)} +...,\;\; j=1,2,...,N,\\
f  &= &1 +\chi^2 f_2 +\chi^4 f_4+...,
\end{eqnarray}
\end{subequations}
where $\chi$ is the formal expansion parameter. The resulting set of 
equations, after collecting the terms with the same power in $\chi$, 
can be solved recursively to obtain the forms of $g^{(j)}$'s and $f$. Though a
formal closed form solution of the $N$-soliton expression of Eq. (1) as a ratio of
two ($N\times N$) determinants can be given [23], it becomes necessary to
deduce the explicit expressions (which is nontrivial) in order to understand the
interaction properties at least for the lower order solitons. In the next
section we will only present the minimum details.
\section{MultiSoliton solutions for $N=2$ case}
As a prelude to understand the nature of soliton solutions for arbitrary 
 $N$-CNLS equations, we first present the bright one- and two- soliton 
solutions of 
Eq. (1) with  $N$=2 (Manakov) case as given in Ref. [15] and then extend the 
analysis to obtain the explicit higher order soliton solutions.
\subsection{One-soliton solution}
After restricting the power series expansion (4) as 
\bear
g^{(j)}=\chi g_1^{(j)}, \;j=1,2,\;\;
f=1+\chi^2 f_2 ,
\eear
and solving the resulting set of linear partial differential 
equations recursively, one can write down the explicit 
one-soliton solution as
\begin{eqnarray}
\left(
\begin{array}{c}
q_1\\
q_2 
\end{array}
\right)
 = 
\left(
\begin{array}{c}
\alpha_1^{(1)}\\
\alpha_1^{(2)}
\end{array}
\right)\frac{e^{\eta_1}}{1+e^{\eta_1+\eta_1^*+R}}\;\;
 = 
\left(
\begin{array}{c}
A_1 \\
A_2
\end{array}
\right)
\frac{k_{1R}e^{i\eta_{1I}}}{\mbox{cosh}\,(\eta_{1R}+\frac{R}{2})},
\end{eqnarray}                                                                                                     	
where $\eta_1=k_1(t+ik_1z)$, $A_j=\frac{\alpha_1^{(j)}}
{\left[\mu\left(|\al_1^{(1)}|^2+|\al_1^{(2)}|^2\right)\right]^{1/2}}$, $j=1,2$, 
and $e^R=\frac{\mu\left(|\al_1^{(1)}|^2+|\al_1^{(2)}|^2\right)}{(k_1+k_1^*)^2}
$.
Note that this one-soliton solution is characterized by three arbitrary complex 
parameters $\alpha_1^{(1)}$,  $\alpha_1^{(2)}$, and $k_1$. Here the 
amplitude of the soliton in the first and second components (modes) are given 
by $k_{1R}A_1$ and $k_{1R}A_2$, respectively, subject to the condition 
$|A_1|^2+|A_2|^2=\frac{1}{\mu}$, while the soliton velocity in both the modes  
is given by $2k_{1I}$. Here $k_{1R}$ and $k_{1I}$ represent the real and
imaginary parts of the complex parameter $k_1$. The quantity 
$\frac{R}{2k_{1R}}=\frac{1}{2k_{1R}}$ $\mbox{log}$$\left[
\frac{\mu\left(|\al_1^{(1)}|^2+|\al_1^{(2)}|^2\right)}{(k_1+k_1^*)^2}
\right]
$ denotes the position of the soliton.
\subsection{ Two-soliton solution}
The two-soliton solution of the integrable 2-CNLS system has been obtained in
Ref. [15] after 
terminating the power series (4) as
\bes 
\bear
g^{(j)}&=& \chi g_1^{(j)}+\chi^3 g_3^{(j)},\;\; j=1,2, \\
f &=& 1+\chi^2 f_2+\chi^4 f_4,
\eear
\ees
and again solving the resultant linear partial differential equations 
recursively . Then the explicit form of the two-soliton solution can be 
written as 
\begin{subequations}
\bear
q_j=\frac{\alpha_1^{(j)}e^{\eta_1}+\alpha_2^{(j)}e^{\eta_2}
+e^{\eta_1+\eta_1^*+\eta_2+\delta_{1j}}+e^{\eta_1+\eta_2+\eta_2^*
+\delta_{2j}}}
{D},\;\;\;j=1,2,
\eear 
where
\begin{eqnarray}
D &=&  1+e^{\eta_1+\eta_1^*+R_1}
+e^{\eta_1+\eta_2^*+\delta_0}
 +e^{\eta_1^*+\eta_2+\delta_0^*}+e^{\eta_2+\eta_2^*+R_2}
+e^{\eta_1+\eta_1^*+\eta_2+\eta_2^*+R_3}.
\end{eqnarray}
In Eqs. (8), we have defined 
\begin{eqnarray}
\eta_i&=&k_i(t+ik_iz),\;\;
e^{\delta_0} = \frac{\kappa_{12}}{k_1+k_2^*},\;\;
e^{R_1} = \frac{\kappa_{11}}{k_1+k_1^*},\;\;\;\;
e^{R_2}=  \frac{\kappa_{22}}{k_2+k_2^*},\nonumber\\
e^{\delta_{1j}}&=&\frac{(k_1-k_2)(\alpha_1^{(j)}\kappa_{21}
-\alpha_2^{(j)}\kappa_{11})}{(k_1+k_1^*)(k_1^*+k_2)},\;\;
e^{\delta_{2j}}=
\frac{(k_2-k_1)(\alpha_2^{(j)}\kappa_{12}-\alpha_1^{(j)}\kappa_{22})}
{(k_2+k_2^*)(k_1+k_2^*)},\nonumber\\
e^{R_3}&=&  \frac{|k_1-k_2|^2}{(k_1+k_1^*)(k_2+k_2^*)|k_1+k_2^*|^2}
 (\kappa_{11}\kappa_{22}-\kappa_{12}\kappa_{21}),
 \eear
\noindent and
\bear
\kappa_{il}= \frac{\mu\sum_{n=1}^2\alpha_i^{(n)}\alpha_l^{(n)*}}
{\left(k_i+k_l^*\right)},\;i,l=1,2.
\end{eqnarray}
\end{subequations}
The above most general bright two-soliton solution is characterized by 
six 
arbitrary complex parameters $k_1$, $k_2$, $\alpha_1^{(j)}$, and 
$\alpha_2^{(j)}$, $j=1,2$, and it corresponds to the collision of two bright 
solitons. 
Note that in Ref. [15], $\delta_{11}$, $\delta_{12}$, $\delta_{21}$, and
$\delta_{22}$ are called as $\delta_1$, $\delta_1'$, $\delta_2$, and $\delta_2'$,
respectively. The redefined quantities $\delta_{ij}$'s, $i,j=1,2$, are now used
for notational simplicity.

\subsection{ Three-soliton solution }
The two-soliton solution itself is very difficult to derive and complicated 
to analyse [15]. So obtaining the three-soliton solution is a more laborious and 
tedious task. However we have successfully obtained the explicit form of 
the bright three-soliton solution also. In 
order to obtain the three-soliton solution of Eq. (1) for the $N=2$ case we 
terminate the power series (4a) and (4b) as 
\begin{subequations}
\bear
g^{(j)}&=& \chi g_1^{(j)}+\chi^3 g_3^{(j)}+\chi^5g_5^{(j)},\\
f &=& 1+\chi^2 f_2+\chi^4 f_4+\chi^6 f_6,\;\;\;j=1,2.
\eear
\end{subequations}
Substitution of (9) into the bilinear Eqs. (3a) and (3b) yields a set of 
linear partial differential 
equations at various powers of $\chi$ .
 The three-soliton solution consistent 
with these equations is 
\begin{subequations}
\bear
q_j&=& \frac{\alpha_1^{(j)}e^{\eta_1}+\alpha_2^{(j)}e^{\eta_2}+\alpha_3^{(j)}
e^{\eta_3}
+e^{\eta_1+\eta_1^*+\eta_2+\delta_{1j}}+e^{\eta_1+\eta_1^*+\eta_3+\delta_{2j}}
+e^{\eta_2+\eta_2^*+\eta_1+\delta_{3j}}}
{D}\nonumber\\
&&+\frac{e^{\eta_2+\eta_2^*+\eta_3+\delta_{4j}}
+e^{\eta_3+\eta_3^*+\eta_1+\delta_{5j}}+e^{\eta_3+\eta_3^*+\eta_2+\delta_{6j}}
+e^{\eta_1^*+\eta_2+\eta_3+\delta_{7j}}
+e^{\eta_1+\eta_2^*+\eta_3+\delta_{8j}}
}{D} \nonumber\\
 &&+\frac{e^{\eta_1+\eta_2+\eta_3^*+\delta_{9j}}
+e^{\eta_1+\eta_1^*+\eta_2+\eta_2^*+\eta_3+\tau_{1j}}
+e^{\eta_1+\eta_1^*+\eta_3+\eta_3^*+\eta_2+\tau_{2j}}}{D}\nonumber\\
&&+\frac{
e^{\eta_2+\eta_2^*+\eta_3+\eta_3^*+\eta_1+\tau_{3j}}}{D},\;\;j=1,2,
\eear
where
\bear
D &=&1+e^{\eta_1+\eta_1^*+R_1}+e^{\eta_2+\eta_2^*+R_2}+e^{\eta_3+\eta_3^*+R_3}
+e^{\eta_1+\eta_2^*+\del_{10}}+e^{\eta_1^*+\eta_2+\del_{10}^*}\nonumber\\
&&e^{\eta_1+\eta_3^*+\del_{20}}
+e^{\eta_1^*+\eta_3+\del_{20}^*}
+e^{\eta_2+\eta_3^*+\del_{30}}
+e^{\eta_2^*+\eta_3+\del_{30}^*}
+e^{\eta_1+\eta_1^*+\eta_2+\eta_2^*+R_4}\nonumber\\
&&+e^{\eta_1+\eta_1^*+\eta_3+\eta_3^*+R_5}
+e^{\eta_2+\eta_2^*+\eta_3+\eta_3^*+R_6}
+e^{\eta_1+\eta_1^*+\eta_2+\eta_3^*+\tau_{10}}
+e^{\eta_1+\eta_1^*+\eta_3+\eta_2^*+\tau_{10}^*}\nonumber\\
&&+e^{\eta_2+\eta_2^*+\eta_1+\eta_3^*+\tau_{20}}
+e^{\eta_2+\eta_2^*+\eta_1^*+\eta_3+\tau_{20}^*}
+e^{\eta_3+\eta_3^*+\eta_1+\eta_2^*+\tau_{30}}
+e^{\eta_3+\eta_3^*+\eta_1^*+\eta_2+\tau_{30}^*}\nonumber\\
&&+e^{\eta_1+\eta_1^*+\eta_2+\eta_2^*+\eta_3+\eta_3^*+R_7}.\\
\mbox{Here}\qquad \qquad \qquad\nonumber\\
\eta_i&=&k_i(t+ik_iz), i=1,2,3,\\
e^{\delta_{1j}}&=&\frac{(k_1-k_2)(\al_1^{(j)}\kappa_{21}-\al_2^{(j)}\kappa_{11}
)}{(k_1+k_1^*)(k_1^*+k_2)},\;\;
e^{\delta_{2j}}=\frac{(k_1-k_3)(\al_1^{(j)}\kappa_{31}-\al_3^{(j)}\kappa_{11}
)}{(k_1+k_1^*)(k_1^*+k_3)},\nonumber\\
e^{\delta_{3j}}&=&\frac{(k_1-k_2)(\al_1^{(j)}\kappa_{22}-\al_2^{(j)}\kappa_{12}
)}{(k_1+k_2^*)(k_2+k_2^*)},\;\;
e^{\delta_{4j}}=\frac{(k_2-k_3)(\al_2^{(j)}\kappa_{32}-\al_3^{(j)}\kappa_{22}
)}{(k_2+k_2^*)(k_2^*+k_3)},\nonumber\\
e^{\delta_{5j}}&=&\frac{(k_1-k_3)(\al_1^{(j)}\kappa_{33}-\al_3^{(j)}\kappa_{13}
)}{(k_3+k_3^*)(k_3^*+k_1)},\;\;
e^{\delta_{6j}}=\frac{(k_2-k_3)(\al_2^{(j)}\kappa_{33}-\al_3^{(j)}\kappa_{23}
)}{(k_3^*+k_2)(k_3^*+k_3)},\nonumber\\
e^{\delta_{7j}}&=&\frac{(k_2-k_3)(\al_2^{(j)}\kappa_{31}-\al_3^{(j)}\kappa_{21}
)}{(k_1^*+k_2)(k_1^*+k_3)},\;\;
e^{\delta_{8j}}=\frac{(k_1-k_3)(\al_1^{(j)}\kappa_{32}-\al_3^{(j)}\kappa_{12}
)}{(k_1+k_2^*)(k_2^*+k_3)},\nonumber\\
e^{\delta_{9j}}&=&\frac{(k_1-k_2)(\al_1^{(j)}\kappa_{23}-\al_2^{(j)}\kappa_{13}
)}{(k_1+k_3^*)(k_2+k_3^*)},\nonumber\\
e^{\tau_{1j}}&=&\frac{(k_2-k_1)(k_3-k_1)(k_3-k_2)(k_2^*-k_1^*)}
{(k_1^*+k_1)(k_1^*+k_2)(k_1^*+k_3)(k_2^*+k_1)(k_2^*+k_2)(k_2^*+k_3)}\nonumber\\
&&\times
\left[\al_1^{(j)}(\kappa_{21}\kappa_{32}-\kappa_{22}\kappa_{31})
+\al_2^{(j)}(\kappa_{12}\kappa_{31}-\kappa_{32}\kappa_{11})
+\al_3^{(j)}(\kappa_{11}\kappa_{22}-\kappa_{12}\kappa_{21})
\right],\nonumber\\
e^{\tau_{2j}}&=&\frac{(k_2-k_1)(k_3-k_1)(k_3-k_2)(k_3^*-k_1^*)}
{(k_1^*+k_1)(k_1^*+k_2)(k_1^*+k_3)(k_3^*+k_1)(k_3^*+k_2)(k_3^*+k_3)}\nonumber\\
&&\times
\left[\al_1^{(j)}(\kappa_{33}\kappa_{21}-\kappa_{31}\kappa_{23})
+\al_2^{(j)}(\kappa_{31}\kappa_{13}-\kappa_{11}\kappa_{33})
+\al_3^{(j)}(\kappa_{23}\kappa_{11}-\kappa_{13}\kappa_{21})
\right],\nonumber\\
e^{\tau_{3j}}&=&\frac{(k_2-k_1)(k_3-k_1)(k_3-k_2)(k_3^*-k_2^*)}
{(k_2^*+k_1)(k_2^*+k_2)(k_2^*+k_3)(k_3^*+k_1)(k_3^*+k_2)(k_3^*+k_3)}\nonumber\\
&&\times
\left[\al_1^{(j)}(\kappa_{22}\kappa_{33}-\kappa_{23}\kappa_{32})
+\al_2^{(j)}(\kappa_{13}\kappa_{32}-\kappa_{33}\kappa_{12})
+\al_3^{(j)}(\kappa_{12}\kappa_{23}-\kappa_{22}\kappa_{13})
\right],\nonumber\\
\eear
\bear
e^{R_m}&=&\frac{\kappa_{mm}}{k_m+k_m^*}, \;\;m=1,2,3,\;\;
e^{\del_{10}}=\frac{\kappa_{12}}{k_1+k_2^*},\;\;
e^{\del_{20}}=\frac{\kappa_{13}}{k_1+k_3^*},\;\;
e^{\del_{30}}=\frac{\kappa_{23}}{k_2+k_3^*},\nonumber\\
e^{R_4}&=&\frac{(k_2-k_1)(k_2^*-k_1^*)}
{(k_1^*+k_1)(k_1^*+k_2)(k_1+k_2^*)(k_2^*+k_2)}
\left[\kappa_{11}\kappa_{22}-\kappa_{12}\kappa_{21}\right],\nonumber\\
e^{R_5}&=&\frac{(k_3-k_1)(k_3^*-k_1^*)}
{(k_1^*+k_1)(k_1^*+k_3)(k_3^*+k_1)(k_3^*+k_3)}
\left[\kappa_{33}\kappa_{11}-\kappa_{13}\kappa_{31}\right],\nonumber\\
e^{R_6}&=&\frac{(k_3-k_2)(k_3^*-k_2^*)}
{(k_2^*+k_2)(k_2^*+k_3)(k_3^*+k_2)(k_3+k_3^*)}
\left[\kappa_{22}\kappa_{33}-\kappa_{23}\kappa_{32}\right],\nonumber\\
e^{\tau_{10}}&=&\frac{(k_2-k_1)(k_3^*-k_1^*)}
{(k_1^*+k_1)(k_1^*+k_2)(k_3^*+k_1)(k_3^*+k_2)}
\left[\kappa_{11}\kappa_{23}-\kappa_{21}\kappa_{13}\right],\nonumber\\
e^{\tau_{20}}&=&\frac{(k_1-k_2)(k_3^*-k_2^*)}
{(k_2^*+k_1)(k_2^*+k_2)(k_3^*+k_1)(k_3^*+k_2)}
\left[\kappa_{22}\kappa_{13}-\kappa_{12}\kappa_{23}\right],\nonumber\\
e^{\tau_{30}}&=&\frac{(k_3-k_1)(k_3^*-k_2^*)}
{(k_2^*+k_1)(k_2^*+k_3)(k_3^*+k_1)(k_3^*+k_3)}
\left[\kappa_{33}\kappa_{12}-\kappa_{13}\kappa_{32}\right],\nonumber\\
e^{R_7}&=& \frac{|k_1-k_2|^2|k_2-k_3|^2|k_3-k_1|^2}
{(k_1+k_1^*)(k_2+k_2^*)(k_3+k_3^*)|k_1+k_2^*|^2|k_2+k_3^*|^2|k_3+k_1^*|^2}
\nonumber\\
&&\times\left[(\kappa_{11}\kappa_{22}\kappa_{33}-
\kappa_{11}\kappa_{23}\kappa_{32})
+(\kappa_{12}\kappa_{23}\kappa_{31}-
\kappa_{12}\kappa_{21}\kappa_{33})\right .\nonumber\\
&&\left.+(\kappa_{21}\kappa_{13}\kappa_{32}-
\kappa_{22}\kappa_{13}\kappa_{31})\right],
\eear
and
\bear
\kappa_{il}= \frac{\mu\sum_{n=1}^2\alpha_i^{(n)}\alpha_l^{(n)*}}
{\left(k_i+k_l^*\right)},\;i,l=1,2,3.
\end{eqnarray} 
\end{subequations}
The above three-soliton solution represents three soliton interaction in the 2-CNLS 
equations and  is characterized by nine arbitrary complex 
parameters $\al_i^{(j)}$'s and $k_i$'s, $i=1,2,3$, $j=1,2$. One can also check that the
above general three-soliton solution of the 2-CNLS equations reduces to that
of the solution given in
Ref. [24] for the particular case $\al_3^{(1)}=1$. Further, the form in which we have
presented the solution eases the complexity in generalizing the solution to
multicomponent case as well as to higher order soliton soutions.
\subsection{Four-soliton solution}
The expression is quite lengthy, but it is written explicitly in terms of
exponential functions so as to check the pair-wise nature of collisions. We indicate the form in Appendix A. One can
generalize these expressions for the arbitrary $N$ case also. However, it is too
complicated to present the explicit form and so we desist from doing so.
\section{Multisoliton solutions for the N-CNLS equations}
As mentioned in the introduction, results are scarce for Eq. (1) with 
$N>2$ and there exists a large class of physical systems in which the  $N$-CNLS equations
occur naturally. Further, in the context of spatial solitons in
photorefractive media, each fundamental soliton 
can be ``spread out" into several incoherent components [25], as defined by the 
polarization vectors. Obtaining one-, two-, and higher order soliton solutions 
of  $N$-CNLS equations will be of considerable significance in these topics. 
In order to 
study the solution properties of such systems  we consider  
the integrable $N$-CNLS equations (1). Following the procedure mentioned in the
earlier section we obtain the one-, two-, and three- (as well as four-) 
soliton solutions of $N$-CNLS
equations as given
below. Particularly the so-called partially coherent solitons will turn out to
be special cases of these soliton solutions ( see Sec. IX below).
\subsection{ One-soliton solution}
The one-soliton solution of Eq. (1) is obtained as
\begin{eqnarray}
\left(
q_1,
q_2,\ldots, 
q_N
\right)^T
& = &
k_{1R}e^{i\eta_{1I}}}{\mbox{sech}\,\left(\eta_{1R}+\frac{R}{2}\right)\left(
A_1,
A_2,\ldots,
A_N
\right)^T
, 
\end{eqnarray}
where 
$\eta_1=k_1(t+ik_1z)$,
$A_j =\alpha_1^{(j)}/\Delta$,
$\Delta = (\mu(\sum_{j=1}^N{|\alpha_1^{(j)}|^2}))^{1/2}$,
$e^R=\Delta^2/(k_1+k_1^*)^2$,
$\alpha_1^{(j)}$ and $k_1$, 
$j$$=$$1,2\ldots,N,$ are $(N+1)$ arbitrary complex parameters. 
Further $k_{1R}A_j$ gives the amplitude of the $j$th mode ($j=1,2,\ldots, N$)
 and $2k_{1I}$ is the soliton velocity in all the $N$ modes.
\subsection{Two-soliton solution}
The two-soliton solution of Eq. (1) can be obtained by following the procedure
given for the 2-component case. It can be written as
\begin{eqnarray}
q_j
&=&
\frac{\al_1^{(j)}e^{\eta_1}+\al_2^{(j)}e^{\eta_2}+
e^{\eta_1+\eta_1^*+\eta_2+\delta_{1j}}
+e^{\eta_1+\eta_2+\eta_2^*+\delta_{2j}}}{D},\;\;j=1,2,...,N,
\end{eqnarray}
where
the denominator $D$ and the co-efficients $e^{R_1}$, $e^{R_2}$, $e^{R_3}$, 
$e^{\delta_0}$, $e^{\delta_0^*}$, $e^{\delta_{1j}}$, and $e^{\delta_{2j}}$ ,
bear the same form as given in (8c) and (8d), except that 
$j$ now runs from 1 to $N$ and
that $\kappa_{il}$'s are redefined as 
\bear
\kappa_{il}=\frac{\mu\sum_{n=1}^N\al_i^{(n)}\al_l^{(n)*}
}
{(k_i+k_l^*)},\;\;i,l=1,2.
\eear
One may also note that the above two-soliton solution depends on $2(N+1)$
arbitrary complex parmeters $\al_1^{(j)}$, $\al_2^{(j)}$, $k_1$, and $k_2$, 
$j=1,2,...,N$.

\subsection{Three-soliton solution}
Following the procedure given in the previous section we obtain the  
3-soliton solution to the $N$-CNLS equations as
\begin{subequations}
\bear
q_j&=& \frac{\alpha_1^{(j)}e^{\eta_1}+\alpha_2^{(j)}e^{\eta_2}+\alpha_3^{(j)}
e^{\eta_3}
+e^{\eta_1+\eta_1^*+\eta_2+\delta_{1j}}+e^{\eta_1+\eta_1^*+\eta_3+\delta_{2j}}
+e^{\eta_2+\eta_2^*+\eta_1+\delta_{3j}}}
{D}\nonumber\\
&&+\frac{e^{\eta_2+\eta_2^*+\eta_3+\delta_{4j}}
+e^{\eta_3+\eta_3^*+\eta_1+\delta_{5j}}+e^{\eta_3+\eta_3^*+\eta_2+\delta_{6j}}
+e^{\eta_1^*+\eta_2+\eta_3+\delta_{7j}}
+e^{\eta_1+\eta_2^*+\eta_3+\delta_{8j}}
}{D} \nonumber\\
 &&+\frac{e^{\eta_1+\eta_2+\eta_3^*+\delta_{9j}}
+e^{\eta_1+\eta_1^*+\eta_2+\eta_2^*+\eta_3+\tau_{1j}}}{D}\nonumber\\
&&+\frac{e^{\eta_1+\eta_1^*+\eta_3+\eta_3^*+\eta_2+\tau_{2j}}
+e^{\eta_2+\eta_2^*+\eta_3+\eta_3^*+\eta_1+\tau_{3j}}}{D},\;\;j=1,2,..., N.
\eear
Here also the denominator $D$ and all the other quantities are the  
same as those given under Eq. (10) except for the redefinition of 
$\kappa_{il}$'s 
as 
\bear
\kappa_{il}=\frac{\mu\sum_{n=1}^N\al_i^{(n)}\al_l^{(n)*}
}
{(k_i+k_l^*)},\;\;i,l=1,2,3.
\eear
\end{subequations}
It can be observed from the above expression that as the number of solitons 
increases the complexity  also increases and the present
three-soliton solution is characterized by $3(N+1)$ complex 
parameters
$\al_1^{(j)}$,  $\al_2^{(j)}$,  $\al_3^{(j)}$, $j=1,2,...,N$,  $k_1$, $k_2$ 
and $k_3$.

The above procedure can be generalized to obtain four-soliton solution and
higher order soliton solutions as discussed in the case of 2-CNLS equations
straightforwardly and one can identify the $N$-soliton solution of $N$-CNLS will be
dependent on $N(N+1)$ arbitrary complex parameters.

\section{Shape changing nature of soliton interactions and intensity
redistributions}
The remarkable fact about the above bright soliton solutions of the 
integrable CNLS system is that they exhibit 
fascinating shape changing (intensity redistribution / energy exchange) 
collisions as we will see below. This interesting 
behaviour has been reported in Ref. [15] for the two-soliton 
solution of the 2-CNLS equations. In a very recent letter [20], the present authors have 
constructed the two-soliton solution of the 3-CNLS and generalized it to  $N$-CNLS, 
for arbitrary  $N$, and briefly indicated similar shape changing collision dynamics of two interacting bright solitons. 
As these $N$-CNLS equations arise in diverse areas of physics as mentioned in the 
introduction, it is of interest to analyse the interaction properties of the 
 soliton solutions of 2-, 3-, and  $N$-CNLS equations. The collision dynamics can be well 
understood by making appropriate asymptotic analysis to the soliton solutions given in the 
previous sections. Such an analysis will then be used to identify suitable
generalized linear fractional transformations in the next section, to obtain 
possible multistate
logic.

\subsection{Asymptotic analysis of two-soliton solution of 2-CNLS equations}
To start with we shall briefly review the collision properties associated with
the two-soliton solution
(8) of the 2-CNLS equations discussed in Ref. [15] in order to extend
the ideas to the $N$-CNLS case. 
Without loss of generality, we
assume that $k_{jR}$$>$$0$ and $k_{1I}$$>$$k_{2I}$, $k_j=k_{jR}+ik_{jI},$ 
$ j=1,2$,
which corresponds to a head-on collision of the solitons (For the case
$k_{1I}=k_{2I}$, see Sec. IX below). For the above
parametric choice, the variables $\eta_{jR}$'s (real part of $\eta_j$) 
for the two solitons behave asymptotically as 
(i)$\eta_{1R}\sim 0$, $\eta_{2R} \rightarrow \pm\infty$ as 
$z \rightarrow \pm \infty$$\;$and$\;$
(ii)$\eta_{2R}\sim 0$, $\eta_{1R} \rightarrow\mp\infty$ as $z 
\rightarrow \pm
\infty.$$\;\;$
This leads to the following asymptotic forms for the two-soliton solution. (For
other choices of $k_{iR}$ and $k_{iI}$, $i=1,2$, 
similar analysis as given below
can be performed straightforwardly).

\noindent\underline{(i) Before Collision (limit $z \rightarrow -\infty$)}

\noindent(a)\underline{\it Soliton 1}
($\eta_{1R} \approx 0$, $\eta_{2R}$$\rightarrow$$ -\infty$):\\  
\begin{subequations}
\begin{eqnarray}
\left(
\begin{array}{c}
q_1\\
q_2 
\end{array}
\right)
\rightarrow
\left(
\begin{array}{c}
A_1^{1-} \\
A_2^{1-}
\end{array}
\right)
k_{1R}e^{i\eta_{1I}}{\mbox{sech}\,\left(\eta_{1R}+\frac{R_1}{2}\right)}, 
\end{eqnarray}
where 
\bear
\left(
\begin{array}{c}
A_1^{1-}\\
A_2^{1-}
\end{array}
\right) & = & 
\left(
\begin{array}{c}
\alpha_1^{(1)} \\
\alpha_1^{(2)}
\end{array}
\right)
\frac{e^{-R_1/2}}{(k_1+k_1^*)}.
\eear
\end{subequations}
The quantity $e^{R_1}$ is defined in Eq. (8c).

\noindent (b) \underline{\it Soliton 2}
($\eta_{2R} \approx 0$, $\eta_{1R}$ $\rightarrow$$ \infty$):
\bes
\begin{eqnarray}
\left(
\begin{array}{c}
q_1\\
q_2 
\end{array}
\right)
 \rightarrow 
\left(
\begin{array}{c}
A_1^{2-} \\
A_2^{2-}
\end{array}
\right)
k_{2R}e^{i\eta_{2I}}{\mbox{sech}\,\left (\eta_{2R}+\frac{(R_3 - R_1)}{2}\right)},
\end{eqnarray} 
where 
\bear
\left(
\begin{array}{c}
A_1^{2-}\\
A_2^{2-}
\end{array}
\right)=
\left(
\begin{array}{c}
 e^{\delta_{11}}\\
e^{\delta_{12}}
\end{array}
\right) \frac{e^{-(R_1+R_3)/2}}{(k_2+k_2^*)}.
\eear 
\end{subequations}
The quantities in the above expression are again defined in Eq. (8c).

\noindent \underline{ (ii) After Collision (limit $z \rightarrow \infty$)}

Similarly, for $z \rightarrow \infty$, we have the following forms for solitons
$S_1$ and $S_2$.

\noindent (a) \underline{\it Soliton 1}
($\eta_{1R} \approx 0$, $\eta_{2R}$ $\rightarrow$ $ \infty$):
\begin{subequations} 
\begin{equation}
\left(
\begin{array}{c}
q_1\\
q_2 
\end{array}
\right)
 \rightarrow 
\left(
\begin{array}{c}
A_1^{1+} \\
A_2^{1+}
\end{array}
\right)
k_{1R}e^{i\eta_{1I}}{\mbox{sech}\,\left (\eta_{1R}+\frac{(R_3 - R_2)}
{2}\right)},
\end{equation} 
where 
\bear
\left(
\begin{array}{c}
A_1^{1+}\\
A_2^{1+}
\end{array}
\right)=
\left(
\begin{array}{c}
 e^{\delta_{21}}\\
e^{\delta_{22}}
\end{array}
\right)
\frac{e^{-(R_2+R_3)/2}}{(k_1+k_1^*)} .
\eear
\end{subequations} 

\noindent (b) \underline{\it Soliton 2}
($\eta_{2R}\approx 0$, $\eta_{1R}$ $\rightarrow$$ -\infty$):
\begin{subequations}
\begin{eqnarray}
\left(
\begin{array}{c}
q_1\\
q_2 
\end{array}
\right)
\rightarrow
\left(
\begin{array}{c}
A_1^{2+} \\
A_2^{2+}
\end{array}
\right)
k_{2R}e^{i\eta_{2I}}{\mbox{sech}\,\left(\eta_{2R}+\frac{R_2}{2}\right)}, 
\end{eqnarray}
where
\bear
\left(
\begin{array}{c}
A_1^{2+}\\
A_2^{2+}
\end{array}
\right) =
\left(
\begin{array}{c}
\alpha_2^{(1)}\\
\alpha_2^{(2)}
\end{array}
\right) 
\frac{e^{-R_2/2}}{(k_2+k_2^*)}.
\eear
\end{subequations}
In the above expressions for $S_1$ and $S_2$ after collision the quantities
$e^{R_2}$, $e^{R_3}$ , $e^{\delta_{21}}$ and $e^{\delta_{22}}$ are defined in
Eq. (8c).
\subsection{Asymptotic analysis of the 2-soliton solution of N-CNLS equations}
We require the asymptotic forms of the 2-soliton
solutions for arbitrary  $N$ case in the following section in order to identify a 
generalized linear fractional transformation for the amplitude redistribution
among the components.
To get the asymptotic forms of 2-soliton solution of the  $N$-CNLS case, as may be
checked by a careful asymptotic analysis along the lines of the $N=2$ case, we simply
increase the number of components in the $A^{\pm}$ vectors above up to $N$ ($A^{\pm} =
(A_1^{\pm}, A_2^{\pm}, ..., A_N^{\pm})^T$) by adding two more complex parameters $\alpha_1^{(i)}$,
$\alpha_2^{(i)}$, $i=3,4,...,N$, to each of the components so that the forms of the
quantities $e^{R_1}$, $e^{R_2}$, $e^{R_3}$, $e^{\delta_{11}}$,
$e^{\delta_{12}}$, $e^{\delta_{21}}$, $e^{\delta_{22}}$ in Eq. (8c) remain
the same as above except for the replacement of the range of the summation in
$\kappa_{il}$ (Eq. (8d)) from $n=1,2$ to $n=1,2,...,N$. As an example, in the
following we give
the asymptotic forms of two-soliton solution of the $N$-CNLS equations with 
$N=3$, for the case $k_{lR}>0$, $l=1,2$, and $k_{1I}>k_{2I}$. For other
possibilities similar analysis can be made.

\noindent\underline{(i) Before Collision (limit $z \rightarrow -\infty $)}

\noindent(a) \underline{\it Soliton 1} 
($\eta_{1R} \approx 0$, $\eta_{2R}$$\rightarrow$$ -\infty$):
\begin{subequations}
\begin{eqnarray}
\left(
\begin{array}{c}
q_1\\
q_2\\
q_3 
\end{array}
\right)
&\approx&
\left(
\begin{array}{c}
A_1^{1-} \\
A_2^{1-}\\
A_3^{1-}
\end{array}
\right)
k_{1R}e^{i\eta_{1I}}{\mbox{sech}\,\left(\eta_{1R}+\frac{R_1}{2}\right)}, 
\end{eqnarray}
where
\bear
\left(
\begin{array}{c}
A_1^{1-}\\
A_2^{1-}\\
A_3^{1-}
\end{array}
\right) = 
\left(
\begin{array}{c}
\alpha_1^{(1)}\\
\alpha_1^{(2)}\\
\alpha_1^{(3)}
\end{array}
\right)\frac{e^{-R_1/2}}{(k_1+k_1^*)}.
\eear
\ees
(b) \underline{\it Soliton 2} ($\eta_{2R} \approx 0$, $\eta_{1R}$ $\rightarrow$$ \infty$):
\bes
\begin{equation}
\left(
\begin{array}{c}
q_1\\
q_2 \\
q_3
\end{array}
\right)
 \approx
\left(
\begin{array}{c}
A_1^{2-} \\
A_2^{2-}\\
A_3^{2-}
\end{array}
\right)
k_{2R}e^{i\eta_{2I}}{\mbox{sech}
\,\left (\eta_{2R}+\frac{(R_3 - R_1)}{2}\right)},
\end{equation} 
where
\bear
\left(
\begin{array}{c}
A_1^{2-} \\
A_2^{2-}\\
A_3^{2-}
\end{array}
\right) 
= 
\left(
\begin{array}{c}
e^{\delta_{11}}\\ 
e^{\delta_{12}}\\
e^{\delta_{13}}
\end{array}
\right)\frac{e^{-(R_1+R_3)/2}}{(k_2+k_2^*)}.
\eear
\ees
\bes
\noindent\underline{(ii) After Collision (limit $z \rightarrow \infty $)}

\noindent (a) \underline{\it Soliton 1} 
($\eta_{1R} \approx 0$, $\eta_{2R}$ $\rightarrow$ $ \infty$):
\begin{equation}
\left(
\begin{array}{c}
q_1\\
q_2 \\
q_3 
\end{array}
\right)
 \approx
\left(
\begin{array}{c}
A_1^{1+} \\
A_2^{1+} \\
A_3^{1+}
\end{array}
\right)
k_{1R}e^{i\eta_{1I}} {\mbox{sech}\,\left(\eta_{1R}+\frac{(R_3 - R_2)}{2}\right)}, 
\end{equation}
where
\bear
\left(
\begin{array}{c}
A_1^{1+} \\
A_2^{1+}\\
A_3^{1+}
\end{array}
\right) 
= 
\left(
\begin{array}{c}
e^{\delta_{21}}\\ 
e^{\delta_{22}}\\
e^{\delta_{23}}
\end{array}
\right)\frac{e^{-(R_2+R_3)/2}}{(k_1+k_1^*)}.
\eear
\ees
\noindent (b) \underline{\it Soliton 2} ($\eta_{2R}\approx 0$, $\eta_{1R}$ $\rightarrow$$ -\infty$):
\bes
\begin{eqnarray}
\left(
\begin{array}{c}
q_1\\
q_2 \\
q_3
\end{array}
\right)
\approx
\left(
\begin{array}{c}
A_1^{2+} \\
A_2^{2+} \\
A_3^{2+}
\end{array}
\right)
k_{2R}e^{i\eta_{2I}}{\mbox{sech}\,\left(\eta_{2R}+\frac{R_2}{2}\right)}, 
\end{eqnarray}
where
\bear
\left(
\begin{array}{c}
A_1^{2+}\\
A_2^{2+}\\
A_3^{2+}
\end{array}
\right) = 
\left(
\begin{array}{c}
\alpha_2^{(1)}\\
\alpha_2^{(2)}\\
\alpha_2^{(3)}
\end{array}
\right)\frac{e^{-R_2/2}}{(k_2+k_2^*)}.
\eear
\end{subequations}
In the above expressions,  the forms of the quantities $e^{R_j}$,
$e^{\delta_{ij}}$, $i=1,2, j=1,2,3,$ can be identified from
Eqs. (12) and (13) with $N=3$.

\subsubsection{Intensity redistribution}

The above analysis clearly shows that due to the interaction between two copropagating
solitons $S_1$ and $S_2$ in an $N$-CNLS system, 
their amplitudes change from $A_j^{1-}k_{1R}$ 
and $A_j^{2-}k_{2R}$ to $A_j^{1+}k_{1R}$ 
and $A_j^{2+}k_{2R}$, $j=1,2,\ldots,N$, respectively. However, 
during the interaction process
the total energy of each of the solitons is conserved, that is
\bear
\sum_{j=1}^N |A_j^{1\pm}|^2=\sum_{j=1}^N |A_j^{2\pm}|^2=\frac{1}{\mu}.
\eear
Note that this is a
consequence of the conservation of $L^2$ norm.  Another 
noticeable observation of this interaction process is that one can
observe from the equation 
of motion(1) itself, that the intensity of each of the modes is 
separately conserved,  that is,
\bear
\int_{-\infty}^{\infty} |q_j|^2 dz =  \mbox{constant} , \;\;j=1,2,\ldots, N.
\eear
The above two equations (23) and (24) ensure that in a two soliton collision
process (as well as in multisoliton collision processes as will be seen later on) 
the total intensity of individual solitons in all the $N$ modes are
conserved along with conservation of intensity of individual modes (even while
allowing an intensity redistribution). This is a striking feature of 
the integrable
nature of the multicomponent CNLS equations (1).
The change in the
amplitude of each of the solitons in the $j$th mode can be obtained by introducing the transition matrix $T_j^l$,
$j=1,2,...,N$, $l=1,2$, such that 
\bes
\bear
A_j^{l+}=T_j^lA_j^{l-}.
\eear
The form of $T_j^l$'s can be obtained from the above asymptotic analysis as 
\bear
T_j^1=\left(\frac{a_2}{a_2^*}\right)\sqrt{\frac{\kappa_{21}}{\kappa_{12}}}
\left[\frac{1-\lambda_2\left(\frac{\alpha_2^{(j)}}{\alpha_1^{(j)}}\right)}
{\sqrt{1-\lambda_1\lambda_2}}\right], \;j=1,2,...,N,
\eear 
where 
\bear
a_2 = (k_2+k_1^*)\left[(k_1-k_2)\sum_{n=1}^N\alpha_1^{(n)}\alpha_2^{(n)*}
\right]^{1/2},
\eear
and 
\bear
T_j^2=-\left(\frac{a_1}{a_1^*}\right)\sqrt{\frac{\kappa_{21}}{\kappa_{12}}}
\left[\frac{\sqrt{1-\lambda_1\lambda_2}}{1-\lambda_1\left(\frac{\alpha_1^{(j)}}
{\alpha_2^{(j)}}\right)}\right], \;j=1,2,...,N,
\eear
in which 
\bear
a_1 = (k_1+k_2^*)\left[(k_1-k_2)\sum_{n=1}^N\alpha_1^{(n)*}\alpha_2^{(n)}
\right]^{1/2}.
\eear
\ees
In the above expressions 
$\lambda_1=\frac{\kappa_{21}}{\kappa_{11}}$ and 
$\lambda_2=\frac{\kappa_{12}}{\kappa_{22}}$, where $\kappa_{il}$'s , $i,l=1,2$,
are defined in Eq. (13). 
Then the intensity exchange in solitons $S_1 $ and $S_2$ 
due to collision can be obtained by taking the absolute square of Eq. (25b) and
(25d), respectively.

The above expressions for the components of the transition matrix implies that
in general there is a redistribution of the intensities in the $N$ modes of both
the solitons after collision. Only for the special case 
\bear
\frac{\al_1^{(1)}}{\al_2^{(1)}}=\frac{\al_1^{(2)}}{\al_2^{(2)}}=...
=\frac{\al_1^{(N)}}{\al_2^{(N)}},
\eear
there occurs the standard
elastic collision. For all other choices of the parameters, shape changing
(intensity redistribution) collision occurs. 

The two conservation relations (23) and (24) allow the intensity
redistribution to take place in  definite ways. In general, for N-CNLS equations
the intensity redistribution in a two soliton collision can occur in $2^N-2$
ways. Denoting E and S as enhancement and suppression respectively, either
complete or partial, of the intensity of corresponding modes, we table below the
possibilities of intensity redistribution for the case $N=2$ and $N=3$. 
\begin{center}
\begin{tabular}{|c|c|c|}
\hline
Case & $q_1$ & $q_2$ \\
\hline
1 & E & S \\
2 & S & E \\
\hline
\end{tabular}
\end{center}
\begin{center}
(a) $N=2$ case
\end{center}
\begin{center}
\begin{tabular}{|c|c|c|c|}
\hline
Case & $q_1$ & $q_2$ & $q_3$\\
\hline
1 & E & S & S\\
2 & S & E & S\\
3 & S & S & E\\
4 & S & E & E\\
5 & E & S & E\\
6 & E & E & S\\
\hline
\end{tabular}
\end{center}
\begin{center}
(b) $N=3$ case
\end{center}
\begin{center}
Table-I. Possible combinations of intensity redistribution among the modes
of soliton $S_1$ in the two soliton collision process.
\end{center}

\noindent For each of the above choices of
$S_1$, the form of $S_2$ is determined by the conserved quantity (24) for the
intensities of the individual modes. 
For
illustrative purpose, we have shown in Fig. 1 and Fig. 2 a few of such
possibilities of intensity switching for the $N=2$ and $N=3$ cases, respectively.
\begin{figure}[h]
\caption{Two distinct possibilities of the shape changing two soliton collision in the
integrable 2-CNLS system. The parameters are chosen as 
(a) $k_1=1+i$, $k_2=2-i$, $\al_1^{(1)}$ $=$ $\al_1^{(2)}=\al_2^{(2)}=1$,
 $\al_2^{(1)}=\frac{39+80i}{89}$; (b) 
$k_1=1+i$, $k_2=2-i$, $\al_1^{(1)}$ $=$ $0.02+0.1i$,
$\al_1^{(2)}=\al_2^{(1)}=\al_2^{(2)}=1$.}
\end{figure}

\begin{figure}[h]
\centerline{}
\caption{Intensity profiles of the three modes of the two-soliton solution 
in a waveguide described by the 3-CNLS (Eq. (1) with $N=3$) showing  different dramatic 
scenarios of the shape changing collision for various choices of parameters.}
\end{figure}

\subsubsection{Phase shifts}

Further, from the asymptotic forms of the solitons $S_1$ and $S_2$, it can be
observed that the phases of solitons $S_1$ and $S_2$ also change during a
collision process and that the
phase shifts are now not only functions of the parameters $k_1$ and $k_2$ but
also dependent on $\al_i^{(j)}$'s, $i=1,2$, $j=1,2,...,N$. The phase
shift suffered by the soliton $S_1$ during collision is 
\bear
\Phi^1=\frac{\left(R_3-R_1-R_2\right)}{2}\;\;\;
=\left(\frac{1}{2}\right)\mbox{log}\left[\frac{
|k_1-k_2|^2(\kappa_{11}\kappa_{22}-\kappa_{12}\kappa_{21})}
{|k_1+k_2^*|^2\kappa_{11}\kappa_{22}}\right],
\eear 
where $\kappa_{il}$'s are defined in Eq. (13).
Similarly the soliton $S_2$ suffers a phase shift 
\bear
\Phi^2&=&-\frac{\left(R_3-R_2-R_1\right)}{2} =-\Phi^1\;. 
\eear
Then the absolute value of phase shift suffered by the two solitons is 
\begin{equation}
|\Phi|=|\Phi^1|=|\Phi^2|.
\end{equation}

Let us consider the case $N=2$. 
For a better understanding let us consider the
pure elastic
collision case 
($\al_1^{(1)}:\al_2^{(1)}=\al_1^{(2)}:\al_2^{(2)}$) corresponding to parallel
modes. Here the absolute phase shift (see Eq. (29)) can be obtained as
\bear 
|\Phi| =\left|\mbox{log}\left[\frac{|k_1-k_2|^2}{|k_1+k_2^*|^2} \right] \right|
=2\left|\mbox{log}\left[\frac{|k_1-k_2|}{|k_1+k_2^*|} \right] \right|. 
\eear
Similarly for the case corresponding to orthogonal modes 
($\al_1^{(1)}:\al_2^{(1)} = \infty$, $\al_1^{(2)}:\al_2^{(2)}=0$) 
the absolute phase shift is found from Eqs. (27)-(29) to be 
\bear 
|\Phi| =\left|\mbox{log}\left[\frac{|k_1-k_2|}{|k_1+k_2^*|} \right] \right|. 
\eear
The absolute value of the phase shift takes intermediate values for other
choices of the parameters $\alpha_i^{(j)}$'s, $i=1,2$, $j=1,2,...,N$. 
Thus phase shifts do vary depending on $\al_i^{(j)}$'s (amplitudes) 
for fixed $k_i$'s. In Fig. 3, we plot the change of $|\Phi|$ as a function of
$\al_1^{(1)}$, when it is real, at $\al_1^{(2)} = \al_2^{(2)}=1$,
$\al_2^{(1)}=\frac{39+80i}{89}$, $k_1=1+i$, and $k_2=2-i$. 
Similar analysis can be done for the $N=3$ case and for the arbitrary $N$ case.
\begin{figure}[h]
\caption{Plot of the magnitude of phase shift as a function of the parameter 
$\al_1^{(1)}$, when it is real (for illustrative purpose), see Eqs. (29-31). The
other parameters are chosen as $k_1=1+i$, $k_2=2-i$, $\al_1^{(2)}=\al_2^{(2)}=1$
and $\al_2^{(1)}=\frac{39+80i}{89}$.}
\end{figure}

\subsubsection {Relative separation distance}

Ultimately the above phase shifts make the relative separation distance 
$t_{12}^{\pm}$ between the
solitons (that is, the position of $S_2$ (at $z \rightarrow \pm\infty$) minus position of
$S_1$  (at $z \rightarrow \pm \infty))$ also to vary during collision, depending
upon the amplitudes of the modes. The change in the relative separation distance is found to
be
\begin{equation}
\Delta t_{12}=t_{12}^--t_{12}^+ =\frac{(k_{1R}+k_{2R})}{k_{1R}k_{2R}} \Phi^1.
\end{equation}
Thus as a whole the intensity profiles of the two solitons in different modes 
as
well as the phases and hence the relative separation distance are nontrivially
dependent on $\al_i^{(j)}$'s and vary
as a result of soliton interaction. 

\section {Generalized Linear Fractional Transformations and Multistate Logic}
The intensity redistribution was characterized by the transition matrix as given
in Eq. (25) in the previous section. Interestingly, this redistribution can
also be viewed as a linear fractional transformation (LFT) as already pointed
out by Jakubowski et al. [18]. However, no systematic derivation of such a
connection was made. In this section, we point out that in fact a reformulation
of Eq. (25) allows one to deduce such a LFT in a systematic way. This in turn
allows us to generalize the procedure to the $N$-component case leading to a
generalized LFT for the amplitude change during soliton collision thereby
leading to a multistate logic. 
\subsection {N=2 case}
For the $N=2$ case, the amplitude change in the two modes of soliton 1 
after interaction given by Eq. (25) can be reexpressed by the following 
transformation, which can be deduced from
comparison of expressions (15b) and (17b): 
\begin{subequations}
\begin{eqnarray}
A_1^{1+} &=& \Gamma C_{11}A_1^{1-} + \Gamma C_{12}A_2^{1-},\nonumber\\
A_2^{1+} &=& \Gamma C_{21}A_1^{1-} + \Gamma C_{22}A_2^{1-}.
\end{eqnarray}
Here 
\bear
\Gamma &=& \Gamma(A_1^{1-},A_2^{1-},A_1^{2-},A_2^{2-})\nonumber\\
&\equiv& \left(\frac{a_2}{a_2^*}\right)\left[\frac{1}
{((\alpha_1^{(1)}\alpha_2^{(1)*}+
\alpha_1^{(2)}\alpha_2^{(2)*})(\alpha_2^{(1)}\alpha_2^{(1)*}+
\alpha_2^{(2)}\alpha_2^{(2)*}))}\right] \left[\frac{1}{|\kappa_{12}|^2}
-\frac{1}{\kappa_{11}\kappa_{22}}\right]^{-1/2},\nonumber\\
\eear 
in which $a_2$ is given in Eq. (25c).
The forms of $C_{ij}$'s, $i,j=1,2$, read
as 
\bear
C_{11} &=&
\alpha_2^{(1)}\alpha_2^{(1)*}(k_1-k_2)+\alpha_2^{(2)}\alpha_2^{(2)*}(k_1+k_2^*),\\
C_{12} &=& -\alpha_2^{(1)}\alpha_2^{(2)*}(k_2+k_2^*),\\ 
C_{21}& = &-\alpha_2^{(2)}\alpha_2^{(1)*}(k_2+k_2^*),\\
C_{22} &=&
\alpha_2^{(1)}\alpha_2^{(1)*}(k_1+k_2^*)+\alpha_2^{(2)}\alpha_2^{(2)*}(k_1-k_2)
.
\eear
\end{subequations} 
Note that the coefficients $C_{ij}$'s are independent of $\alpha_1^{(j)}$'s and so of
$A_1^{1-}$ and $A_2^{1-}$, that is the $\alpha$-parameters of soliton 1. 
Then from Eqs. (33a) the ratios of the $A_i^{j\pm}$'s, $i,j=1,2,$ can be connected through
 a linear fractional 
transformation(LFT). For example, for soliton 1, from (33a),
\begin{equation}
\rho_{1,2}^{1+}=\frac{A_1^{1+}}{A_2^{1+}}=\frac{C_{11}\rho_{1,2}^{1-}+C_{12}}
{C_{21}\rho_{1,2}^{1-}+C_{22}},
\end{equation}
where $\rho_{1,2}^{1-}$$=$$\frac{A_1^{1-}}{A_2^{1-}}$, in which the superscripts
represent the underlying soliton and the subscripts represent the corresponding
modes. The quantities 
$\rho_{1,2}^{1+}$, 
$\rho_{1,2}^{1-}$, $C_{11}$,  $C_{12}$, $C_{21}$, $C_{22}$, in Eq. (34) are same as the
quantities $\rho_R$, $\rho_1$, $\left(\frac{1-h^*}{\rho_L^*}+\rho_L\right)$,
$h^*\frac{\rho_L}{\rho_L^*}$, $h^*$ and
$\left((1-h^*)\rho_L+\frac{1}{\rho_L^*}\right)$,
respectively, given by Eq. (9) in Ref. [18] in an adhoc way. 
Thus the state of $S_1$ before
and after interaction is characterized by $\rho_{1,2}^{1-}$ and $\rho_{1,2}^{1+}$, respectively.
It is to be noticed that during collision $k_i$'s, $i=1,2$, are unaltered. The 
LFT has 
been profitably used in
Ref. [19] to construct logic gates, associated with the binary logic $\rho =
[0,1]$.
Similar analysis can be done for the soliton 2 also. 
\subsection{N=3 case}
Extending the above analysis, straightforwardly one can relate the 
$A_j^{1\pm}$'s, $j=1,2,3$, for
soliton 1, from Eqs. (19b) and (21b), as
\begin{subequations}
\begin{eqnarray}
A_1^{1+} &=& \Gamma C_{11}A_1^{1-} + \Gamma C_{12}A_2^{1-}
+\Gamma C_{13}A_3^{1-},\\
A_2^{1+} &=& \Gamma C_{21}A_1^{1-} + \Gamma C_{22}A_2^{1-}+
\Gamma C_{23}A_3^{1-},\\
A_3^{1+} &=& \Gamma C_{31}A_1^{1-} + \Gamma C_{32}A_2^{1-}+
\Gamma C_{33}A_3^{1-},
\end{eqnarray}
where 
\bear
\Gamma &=& \Gamma(A_1^{1-},A_2^{1-},A_3^{1-},A_1^{2-},A_2^{2-}, A_3^{2-})\nonumber\\
&\equiv& \left(\frac{a_2}{a_2^*}\right)\left[\frac{1}
{((\alpha_1^{(1)}\alpha_2^{(1)*}+
\alpha_1^{(2)}\alpha_2^{(2)*}+\alpha_1^{(3)}\alpha_2^{(3)*})
(\alpha_2^{(1)}\alpha_2^{(1)*}+
\alpha_2^{(2)}\alpha_2^{(2)*}+\alpha_2^{(3)}\alpha_2^{(3)*}))}\right]\nonumber\\
&&\times \left[\frac{1}{|\kappa_{12}|^2}-\frac{1}{\kappa_{11}\kappa_{22}}\right]^{-1/2},
\eear 
in which $a_2$'s are redefined as 
\bear
a_2=(k_2+k_1^*)[(k_1-k_2)(\alpha_1^{(1)}\alpha_2^{(1)*}+
\alpha_1^{(2)}\alpha_2^{(2)*}+\alpha_1^{(3)}\alpha_2^{(3)*})]^{1/2},
\eear
\end{subequations}
and $\kappa_{il}$'s can be  written from Eq. (13) with $N=3$.
Note that the form of $\Gamma$ is a straightforward extension of the $N=2$ case.
In the above equations the coefficients $C_{ij}$'s, $i,j=1,2,3$, for the 3-CNLS
case can be written down straightforwardly by generalizing the expressions 
(33) corresponding to the two-soliton solution of the two component case.

Thus in the two soliton collision process of the $N=3$ case, for
soliton 1 we obtain the
generalized M{\"o}bius transformation,
\begin{subequations}
\bear
\rho_{1,3}^{1+} &=&
\frac{A_1^{1+}}{A_3^{1+}}= \frac{C_{11}\rho_{1,3}^{1-}+C_{12}\rho_{2,3}^{1-}+C_{13}}
{C_{31}\rho_{1,3}^{1-}+C_{32}\rho_{2,3}^{1-}+C_{33}},\\
\rho_{2,3}^{1+} &=& 
\frac{A_2^{1+}}{A_3^{1+}}=\frac{C_{21}\rho_{1,3}^{1-}+C_{22}\rho_{2,3}^{1-}+C_{23}}
{C_{31}\rho_{1,3}^{1-}+C_{32}\rho_{2,3}^{1-}+C_{33}},
\eear
\end{subequations}
where $\rho_{1,3}^{1-}=\frac{A_1^{1-}}{A_3^{1-}}$ and 
$\rho_{2,3}^{1-}=\frac{A_2^{1-}}{A_3^{1-}}$. 
Similar relations can be obtained for the soliton 2 also. 
\subsection{Arbitrary N case}
Proceeding in a similar fashion one
can construct for the soliton $S_1$ a generalized linear fractional transformation for the N-component case
also which relates the $\rho$ vectors before and after collision.
\bes
\bear
\rho_{i, N}^{1+}=
\frac{A_i^{1+}}{A_N^{1+}}=
\frac{\sum_{j=1}^NC_{ij}\rho_{j, N}^{1-}}
{\sum_{j=1}^NC_{Nj}\rho_{j, N}^{1-}},
\eear
with the condition
\bear
\rho_{NN}^{1-}=1.
\eear
\ees
Here $\rho_{i,N}^{1-}=\frac{A_i^{1-}}{A_N^{1-}}$. Similar expression can be obtained for soliton 2 also. 

The above generalization paves way not only for writing down the bilinear
transformation but also to identify multistate logic. For example, in the
$N=3$ case, the following states are possible: $\rho =[\rho_1,\rho_2]$ 
 $\equiv$ $\left[(0,0), (0,1), (1,0), (1,1) \right]$, where the logical `0'
 state can stand for the complex valued $\rho$ state corresponding to a
 suppression of the amplitude in that mode, while the logical `1' state may
 correspond to enhancement (including no change),
which can be used to perform logical operations, whereas in
the $N=2$ case we have only the two state logic, $\rho =  [0, 1]$. 
This shows that for $N>2$, we will get multistate logic and
we believe that such states can be of distinct advantage in computation. This
kind of study is in progress.

\section{Higher order soliton solutions and their interactions}
Now it is of interest to study the nature of multisoliton collisions making use
of the explicit forms of multisoliton solutions given in Secs. III and IV. Due 
to the complicated nature of the above
bright soliton expressions, it becomes nontrivial to identify the nature of
the collision process. In his paper [13], Manakov pointed out that in general an
N-soliton collision does not reduce to a pair collision due to the nontrivial
dependence of the amplitude of a particular soliton before interaction on the
other soliton parameters. In this section by a careful asymptotic analysis of
the 3-soliton solution (10) of the 2-CNLS equations, which can be deduced to the
$N$-CNLS case without any difficulty, we explicitly demonstrate 
that the
collision process indeed can be considered to occur pair-wise and thereby making
Manakov's statement in proper perspective and clearer.  One can carry 
out a similar analysis for the four-soliton solution given in Appendix A,
generalizing which one can show that
in the higher order solitons of CNLS equations also the collision is pair-wise. 
Such an analysis also reveals the many possibilities for energy exchange among
the modes of the solitons, including the exciting possibility of state
restoration in higher order soliton solutions, a precursor to the construction
of logic gates.

\subsection {Asymptotic analysis of 3-soliton solution of 2-CNLS equations}
Considering the explicit three soliton expression (10), 
without loss of generality, we assume that the 
quantities $k_{1R}$,
$k_{2R}$, and $k_{3R}$ are positive and  $k_{1I}>k_{2I}>k_{3I}$ (For
the equal sign cases $k_{1I}=k_{2I}=k_{3I}$,  see Sec. IX below).
One can carry out similar analysis for other possibilities of $k_{iI}$'s ,
$i=1,2,3,$ also as 
discussed below.
Then for the above condition the variables $\eta_{iR}$'s, $i=1,2,3,$ 
for the three solitons ($S_1$, $S_2$, and $S_3$) take the following 
values asymptotically: 

(i) $\eta_{1R} \approx 0$,  $\eta_{2R} \rightarrow \pm \infty$,
$\eta_{3R} \rightarrow \pm \infty$, as $z \rightarrow \pm \infty$,

(ii) $\eta_{2R} \approx 0$,  $\eta_{1R} \rightarrow \mp \infty$,
$\eta_{3R} \rightarrow \pm \infty$, as $z \rightarrow \pm \infty$,

(iii) $\eta_{3R} \approx 0$,  $\eta_{1R} \rightarrow \mp \infty$,
$\eta_{2R} \rightarrow \mp \infty$, as $z \rightarrow \pm \infty$.

Defining the various quantities $R_i$'s, $i=1,2,...,7$ , $\delta_{lj}$'s,
$l=1,2,...,9$, $j=1,2$, $\tau_{mj}$'s, and  $\tau_{m0}$'s, $m=1,2,3$, as in Eq. (10) 
we have the following limiting forms of the three-soliton solution Eq. (10).

\noindent\underline{(i)Before Collision (limit $ z \rightarrow -\infty$)} 

\noindent(a) \underline{\it Soliton 1}  ($\eta_{1R} \approx 0$, $\eta_{2R} \rightarrow -\infty$,
$\eta_{3R} \rightarrow -\infty$):
\begin{subequations}
\begin{eqnarray}
\left(
\begin{array}{c}
q_1\\
q_2 
\end{array}
\right)
&\approx &
\left(
\begin{array}{c}
A_1^{1-} \\
A_2^{1-}
\end{array}
\right)k_{1R}
{\mbox{sech}\,\left(\eta_{1R}+\frac{R_1}{2}\right)}e^{i\eta_{1I}} ,\\
\left(
\begin{array}{c}
A_1^{1-} \\
A_2^{1-}
\end{array}
\right)
&=&
\left(
\begin{array}{c}
\alpha_1^{(1)}\\
\alpha_1^{(2)}
\end{array}
\right) \frac{e^{\frac{-R_1}{2}}}{(k_1+k_1^*)}.
\eear
\ees
\noindent(b) \underline{\it Soliton 2}  ($\eta_{2R} \approx 0$, $\eta_{1R} \rightarrow \infty$,
$\eta_{3R} \rightarrow -\infty$):
\bes
\begin{eqnarray}
\left(
\begin{array}{c}
q_1\\
q_2 
\end{array}
\right)
&\approx&
\left(
\begin{array}{c}
A_1^{2-} \\
A_2^{2-}
\end{array}
\right)k_{2R}
{\mbox{sech}\,\left(\eta_{2R}+\frac{R_4-R_1}{2}\right)}e^{i\eta_{2I}} ,\\
\left(
\begin{array}{c}
A_1^{2-}\\
A_2^{2-}
\end{array}
\right)
&=&
\left(
\begin{array}{c}
e^{\delta_{11}}\\
e^{\delta_{12}}
\end{array}
\right)\frac{e^{-\frac{(R_1+R_4)}{2}}}{(k_2+k_2^*)}.
\eear
\ees

\noindent (c) \underline{\it Soliton 3}  ($\eta_{3R} \approx 0$, $\eta_{1R} \rightarrow \infty$,
$\eta_{2R} \rightarrow \infty$):
\bes
\begin{eqnarray}
\left(
\begin{array}{c}
q_1\\
q_2 
\end{array}
\right)
&\approx&
\left(
\begin{array}{c}
A_1^{3-} \\
A_2^{3-}
\end{array}
\right)k_{3R}
{\mbox{sech}\,\left(\eta_{3R}+\frac{R_7-R_4}{2}\right)}e^{i\eta_{3I}} ,\\
\left(
\begin{array}{c}
A_1^{3-}\\
A_2^{3-}
\end{array}
\right)
&=&
\left(
\begin{array}{c}
e^{\tau_{11}}\\
e^{\tau_{12}}
\end{array}
\right)\frac{e^{-\frac{(R_4+R_7)}{2}}}{(k_3+k_3^*)}.
\eear
\end{subequations}

\noindent \underline{(ii)After Collision (limit $ z \rightarrow +\infty$)} 

\noindent (a) \underline{\it Soliton 1}  ($\eta_{1R} \approx 0$, $\eta_{2R} \rightarrow \infty$,
$\eta_{3R} \rightarrow \infty$):
\begin{subequations}
\begin{eqnarray}
\left(
\begin{array}{c}
q_1\\
q_2 
\end{array}
\right)
&\approx&
\left(
\begin{array}{c}
A_1^{1+} \\
A_2^{1+}
\end{array}
\right)k_{1R}
{\mbox{sech}\,\left(\eta_{1R}+\frac{R_7-R_6}{2}\right)}e^{i\eta_{1I}} ,\\
\left(
\begin{array}{c}
A_1^{1+}\\
A_2^{1+}
\end{array}
\right)
&=&
\left(
\begin{array}{c}
e^{\tau_{31}}\\
e^{\tau_{32}}
\end{array}
\right)\frac{e^{-\frac{(R_6+R_7)}{2}}}{(k_1+k_1^*)} .
\eear
\ees

\noindent(b) \underline{\it Soliton 2} ($\eta_{2R} \approx 0$, $\eta_{1R} \rightarrow -\infty$,
$\eta_{3R} \rightarrow \infty$):
\bes
\begin{eqnarray}
\left(
\begin{array}{c}
q_1\\
q_2 
\end{array}
\right)
&\approx&
\left(
\begin{array}{c}
A_1^{2+} \\
A_2^{2+}
\end{array}
\right)k_{2R}
{\mbox{sech}\,\left(\eta_{2R}+\frac{R_6-R_3}{2}\right)}e^{i\eta_{2I}} ,\\
\left(
\begin{array}{c}
A_1^{2+}\\
A_2^{2+}
\end{array}
\right)
&=&
\left(
\begin{array}{c}
e^{\delta_{61}}\\
e^{\delta_{62}}
\end{array}
\right)\frac{e^{-\frac{(R_3+R_6)}{2}}}{(k_2+k_2^*)}.
\eear
\ees

\noindent (c) \underline{\it Soliton 3}  ($\eta_{3R} \approx 0$, $\eta_{1R} \rightarrow -\infty$,
$\eta_{2R} \rightarrow -\infty$):
\bes
\begin{eqnarray}
\left(
\begin{array}{c}
q_1^{3+}\\
q_2^{3+} 
\end{array}
\right)
&\approx &
\left(
\begin{array}{c}
A_1^{3+} \\
A_2^{3+}
\end{array}
\right)k_{3R}
{\mbox{sech}\,\left(\eta_{3R}+\frac{R_3}{2}\right)}e^{i\eta_{3I}} ,\\
\left(
\begin{array}{c}
A_1^{3+} \\
A_2^{3+}
\end{array}
\right)
&=&
\left(
\begin{array}{c}
\alpha_3^{(1)}\\
\alpha_3^{(2)}
\end{array}
\right)\frac{e^{-\frac{R_3}{2}}}{(k_3+k_3^*)}.
\eear
\end{subequations}
\subsection{Transition elements}
The above analysis clearly shows that during the three soliton 
interaction process, there is a redistribution of 
intensities among these solitons in the two modes along with amplitude 
dependent phase shifts as in the case of the two soliton interaction. 
The amplitude changes can be expressed in terms of a transition matrix 
$T_j^l$ as
\bear
A_j^{l+}=T_j^lA_j^{l-}, j=1,2, \; l=1,2,3.
\eear
Explicit forms of the entries of the transition matrix quantifying the amount of intensity redistribution
for the three solitons are as follows.

\noindent \underline{Soliton 1:}
\bes
\bear
\left(
\begin{array}{c}
T_1^1\\
T_2^1
\end{array}
\right)
=
\left(
\begin{array}{c}
\frac{e^{\tau_{31}}}{\al_1^{(1)}}\\
\frac{e^{\tau_{32}}}{\al_1^{(2)}}
\end{array}
\right)e^\frac{-(R_6+R_7-R_1)}{2}.
\eear
\underline{Soliton 2:}
\bear
\left(
\begin{array}{c}
T_1^2\\
T_2^2
\end{array}
\right)
=
\left(
\begin{array}{c}
e^{\delta_{61}-\delta_{11}}\\
e^{\delta_{62}-\delta_{12}}
\end{array}
\right)e^\frac{-(R_3+R_6-R_1-R_4)}{2}.
\eear
\underline{Soliton 3:}
\bear
\left(
\begin{array}{c}
T_1^3\\
T_2^3
\end{array}
\right)
=
\left(
\begin{array}{c}
\al_3^{(1)}e^{-\tau_{11}}\\
\al_3^{(2)}e^{-\tau_{12}}
\end{array}
\right)e^\frac{-(R_3-R_4-R_7)}{2}.
\eear
\ees
The various quantities found in the above equations are defined in Eq.
(10). 
\subsection{Phase shifts}
Now let us look into the phase shifts suffered by each of the solitons during collision.
These can be written as 
\begin{subequations}
\bear
S_1 : \Phi^1 &=& \frac{R_7-R_6-R_1}{2},\\
S_2 :\Phi^2 &=& \frac{R_6-R_3-R_4+R_1}{2},\\
S_3 :\Phi^3 &=& \frac{R_3-R_7+R_4}{2}.
\eear
\end{subequations}
Here the quantities $R_1$, $R_2$,..., $R_7$ are as given in Eq. (10). 
Note that each of the phase shifts $\Phi^1$, $\Phi^2$, and $\Phi^3$ contains a
part which depends purely on $k_i$'s, $i=1,2,3$, and another part which depends on
the amplitude (polarization) parameters $\al_i^{(j)}$'s along with $k_i$'s. 

\subsection{Relative separation distances}
As a consequence of the above amplitude dependent phase shifts, the relative
separation distances between the solitons $t_{ij}^{\pm}$ (position of $S_j$
(at $z$ $\rightarrow$ 
$\pm$ $\infty$) - position of $S_i$ (at $z$ $\rightarrow$ $\pm$ $\infty$), 
$i\neq j$, $i<j$, 
$i,j=1,2,3$) also varies as a function of amplitude parameters. The change in
the relative separation distances ($\Delta t_{ij}$ $=$ $t_{ij}^--t_{ij}^+$) 
can be obtained from the asymptotic expressions
(38-43). They are found to be 
\bes
\bear
\Delta t_{12} &=& \frac{\Phi^1 k_{2R}-\Phi^2 k_{1R}}{k_{1R}k_{2R}},\\
\Delta t_{13} &=& \frac{\Phi^1 k_{3R}-\Phi^3 k_{1R}}{k_{1R}k_{3R}},\\
\Delta t_{23} &=& \frac{\Phi^2 k_{3R}-\Phi^3 k_{2R}}{k_{2R}k_{3R}},
\eear
\ees
where $\Phi^j$'s, $j=1,2,3$, are defined in Eq. (46) and $k_{jR}$'s represent
the real parts of $k_j$'s.

\subsection{Nature of Collision}
Now it is of interest to look into the nature of the collisions in the three 
soliton interaction process, that is, whether it is
pair-wise or not. This can be answered from the asymptotic expressions
presented in Eqs. (38) to (46).
For example, let us consider soliton 1 ($S_1$). The net change in the amplitudes of
the two modes of soliton $S_1$ is given by the transition amplitudes $T_i^1$, 
$i=1,2$, that is,
\bear
\left(
\begin{array}{c}
A_1^{1+}\\
A_2^{1+}
\end{array}
\right)
=
\left(
\begin{array}{cc}
T_1^1 & 0\\
0 & T_2^1
\end{array}
\right)
\left(
\begin{array}{c}
A_1^{1-}\\
A_2^{1-}
\end{array}
\right),
\eear
where $T_1^1$ and $T_2^1$ are defined in Eq. (45a).
The above form of transition relations is obtained by expanding (44). 

Let us presume first that 
the collision process is a pair-wise one and then verify this assertion. 
According to our assumption $k_{1I}$ 
$>$ $k_{2I}$ $>$ $k_{3I}$, and so  the first collision occurs between $S_1$ and $S_2$
as shown schematically in Fig. 4. 
\begin{figure}[h]
\caption{A schematic three soliton collision process (for the choice $k_{1R}, k_{2R},
k_{3R} > 0$, $k_{1I}>k_{2I}>k_{3I}$). The effect of phase shifts are not included
in the figure.}
\end{figure}
Then during
collision with $S_2$, the two modes of $S_1$ change their amplitudes
(intensities) by $\tilde T_1^1$ and $\tilde T_2^1$, respectively. Their forms
follow from the basic two soliton interaction process discussed in Sec. V, 
Eq. (25b). This can be
expressed in mathematical form as 
\begin{subequations}
\bear
\left(
\begin{array}{c}
\tilde{A}_1^{1+}\\
\tilde{A}_2^{1+}
\end{array}
\right)
=
\left(
\begin{array}{cc}
\tilde T_1^1 & 0\\
0 & \tilde T_2^1
\end{array}
\right)
\left(
\begin{array}{c}
A_1^{1-}\\
A_2^{1-}
\end{array}
\right),
\eear
where
\bear
\left(
\begin{array}{c}
\tilde T_1^1 \\
\tilde T_2^1
\end{array}
\right)
=
\left(
\begin{array}{c}
\frac{e^{\delta_{31}}}{\al_1^{(1)}}\\
\frac{e^{\delta_{32}}}{\al_1^{(2)}}
\end{array}
\right)e^{\frac{-(R_4+R_2-R_1)}{2}}.
\eear
\end{subequations}
Again the above expressions can be obtained straightforwardly from Eq. (25b)
with $N=2$.

Now the resulting soliton  ($\tilde S_1$), after the first collision, is allowed to collide with
the third soliton ($S_3$) (see Fig. 4). From the asymptotic 
expressions (38-45) and using the above
Eqs. (49), it can be shown that
\begin{subequations}
\bear
\left(
\begin{array}{c}
A_1^{1+}\\
A_2^{1+}
\end{array}
\right)
=
\left(
\begin{array}{cc}
\hat T_1^1 & 0\\
0 & \hat T_2^1
\end{array}
\right)
\left(
\begin{array}{c}
\tilde{A}_1^{1+}\\
\tilde{A}_2^{1+}
\end{array}
\right),
\eear
where
\bear
\left(
\begin{array}{c}
\hat T_1^1 \\
\hat T_2^1
\end{array}
\right)
=
\left(
\begin{array}{c}
e^{\tau_{31}-\delta_{31}}\\
e^{\tau_{32}-\delta_{32}}
\end{array}
\right)e^{\frac{-(R_6+R_7-R_4-R_2)}{2}}.
\eear
\end{subequations}
However using Eq. (49) in (50a), we can write 
\begin{subequations}
\bear
\left(
\begin{array}{c}
A_1^{1+}\\
A_2^{1+}
\end{array}
\right)
&=&
\left(
\begin{array}{cc}
\hat T_1^1 & 0\\
0 & \hat T_2^1
\end{array}
\right)
\left(
\begin{array}{cc}
\tilde T_1^1 & 0\\
0 & \tilde T_2^1
\end{array}
\right)
\left(
\begin{array}{c}
{A}_1^{1-}\\
{A}_2^{1-}
\end{array}
\right)\\
&=&
\left(
\begin{array}{cc}
\hat T_1^1 \tilde T_1^1 & 0\\
0 & \hat T_2^1 \tilde T_2^1
\end{array}
\right)
\left(
\begin{array}{c}
A_1^{1-}\\
A_2^{1-}
\end{array}
\right).
\eear
If this is the collision scenario, then the right hand sides of Eq. (48) and
Eq. (51b)
should be the same, that is, 
\bear
T_1^1&=&\hat T_1^1 \tilde T_1^1 ,  \\
 T_2^1&=&\hat T_2^1 \tilde T_2^1 . 
\eear
\end{subequations}
This can be easily verified to be true directly from the expressions (45) and 
(49-50). In a similar fashion, for
the other two solitons also the transition matrix can be shown as a product of
two matrices corresponding to two collisions, respectively.

Now let us look at the phase shifts. It is also of necessary to identify whether
the total phase shift acquired by each soliton during the three soliton
collision process is a result of two consecutive pair-wise collisions or not. In
this regard, we again focus our attention on soliton 1 ($S_1$) first. 
Let us assume the
collision to be pair-wise.  Then one can
write the phase shift suffered by $S_1$ during the collision based on the
analysis of the two soliton collision process. Following Eq. (27) (with
appropriately changed notations) we can write
the expression for the phase shift suffered by $S_1$ on its collision with $S_2$
as  
\bear
\tilde{\delta}=\frac{R_4-R_2-R_1}{2}.
\eear
Now the outcoming form of $S_1$ (which is $\hat S_1$) is allowed to interact 
with $S_3$ (see Fig. 4). The
phase shift during this second collision can again be found from the
asymptotic expressions (38-43) as 
\bear
\hat{\delta}=\frac{R_7-R_6-R_4+R_2}{2}.
\eear
On the other hand, from the asymptotic expressions (46), the total phase shift suffered by $S_1$ in a
three soliton collision process can be written as 
\bear
\delta &=& \frac{R_7-R_6-R_1}{2}\\
&=& \tilde{\delta}+\hat{\delta}.
\eear
Thus the total phase shift suffered by the soliton 1 is the sum of the phase
shifts suffered by it during  pair-wise collisions with soliton 2 and soliton 3, 
respectively. Similar conclusions can also be drawn on the phase shifts suffered
by the other two solitons as well. Thus the above analysis on the changes in the amplitudes and 
phase shifts
during the three soliton collision process establishes the fact that the 
collisions indeed occur pair-wise.

It may be noted that the above results also imply that the three soliton
collision process is associative and independent of the sequence in which
collisions occur, that is whether the collision occurs in the order $S_1$ $\rightarrow$ 
$S_2$ $\rightarrow$ $S_3$ or $S_1$ $\rightarrow$ 
$S_3$ $\rightarrow$ $S_2$. This property has been anticipated in the numerical
study of Lewis et al. [26], which is now rigorously proved here.
\subsection{Intensity redistributions and shape restoration}
The asymptotic analysis not only explains the nature of the collision
process, but also characterizes the collision process. It is clear from the
above analysis of the three-soliton solution that in general there is an intensity
redistribution among the three solitons due to pair-wise interaction in all the two modes along with
amplitude dependent phase shifts as in the two soliton interaction, subject to
conservation laws. We have
analysed the various three soliton collision scenarios below.

\subsubsection{Elastic collision}

The standard elastic collision property of solitons results for the special
case $\alpha_1^{(1)}:\alpha_2^{(1)}:\alpha_3^{(1)}$ $=$ 
$\alpha_1^{(2)}:\alpha_2^{(2)}:\alpha_3^{(2)}$. The magnitude of the 
transition elements
$|T_j^l|$, $j=1,2$, and $l=1,2,3$, becomes one for this choice of parameters and
there occurs no intensity redistribution among the modes except for phase shifts.
This is shown in Fig. 5 for the 
parametric choice $\alpha_l^{(j)}=1$, $l=1,2,3$, $j=1,2$, $k_1=1+i$, 
$k_2=1.5-0.5 i$, and $k_3 =2.0- i$.
\begin{figure}
\caption{Intensity profiles $|q_1|^2$ and $|q_2|^2$ of the two modes of the 
three-soliton 
solution of the 2-CNLS equations, representing elastic collision, with the
parameters chosen as $k_1=1+i$, $k_2 = 1.5-0.5i$, $k_3 = 2-i$, 
$\al_1^{(1)}= \al_2^{(1)}= \al_3^{(1)}= \al_1^{(2)}=
\al_2^{(2)}=\al_3^{(2)}=1$. }
\end{figure}

\subsubsection{Shape changing (intensity redistribution) collision}

For general  values of the parameters $\al_l^{(j)}$'s, there occurs 
shape changing collisions among the
three solitons, however, leaving the total intensity of each of the solitons 
conserved, that is, 
$|A_1^{l\pm}|^2+|A_2^{l\pm}|^2=\frac{1}{\mu}$, $l=1,2,3$. This intensity
redistribution is accompanied by amplitude dependent phase shifts and changes
in the relative separation distances of the solitons as discussed above. They 
can be calculated
from the expressions (38-47). One such shape changing interaction is depicted in 
Fig. 6 for illustrative purpose. 
The parameters chosen are $k_1=1+i$, $k_2 = 1.5-0.5i$, $k_3 = 2-i$, 
$\al_1^{(1)}=\frac{39-80i}{89}$, $\al_2^{(1)}=\frac{39+80i}{89}$, 
$\al_3^{(1)}=0.3+0.2i$, $\al_1^{(2)}=0.39$, $\al_2^{(2)}=\al_3^{(2)}=1$ .
 In this figure we have shown the 
scenario in which the three solitons in the two modes have different 
amplitudes (intensities) after interaction when compared to the case   
before interaction. Here $S_1$ is allowed to interact with $S_2$
first and then with $S_3$. Due to this collision, in $q_1$ mode the 
intensity of
$S_1$ is suppressed while that of $S_2$ is enhanced along with suppression of
intensity in
$S_3$. On the other hand, the reverse scenario occurs in the $q_2$ mode 
for the three
solitons $S_1$, $S_2$, and $S_3$.
\begin{figure}
\caption{Intensity profiles  $|q_1|^2$ and $|q_2|^2$ of the two modes of the 
three-soliton 
solution of the 2-CNLS equations, representing the shape changing 
(intensity redistribution) collision process for the choice of the parameters, 
$k_1=1+i$, $k_2 = 1.5-0.5i$, $k_3 = 2-i$, 
$\al_1^{(1)}=\frac{39-80i}{89}$, $\al_2^{(1)}=\frac{39+80i}{89}$, 
$\al_3^{(1)}=0.3+0.2i$, $\al_1^{(2)}=0.39$, $\al_2^{(2)}=\al_3^{(2)}=1$.}
\end{figure}

\subsubsection{Shape restoration of any one of the three solitons}

The asymptotic analysis also shows that there is a possibility for any one of the
three solitons to restore its shape (amplitude/intensity) during collision. In this connection, let
us look into how the shape restoring property of $S_1$ occurs during its
collision with the other two solitons (say $S_2$ and $S_3$). We have already shown
that the collision process is a pair-wise one. 
Then the three soliton collision process is equivalent to two pair-wise
collisions. Let the first collision be parametrized by the parameters 
$\al_1^{(1)}$, $\al_1^{(2)}$, $\al_2^{(1)}$, $\al_2^{(2)}$, $k_1$, and 
$k_2$. Now we exploit the arbitrariness involved in choosing the parameters 
$\al_3^{(1)}$ and $\al_3^{(2)}$ in the second collision process in order to make
the net transition amplitude of $S_1$ to be unity, leaving the other two transition
amplitudes of $S_2$ and $S_3$ to vary, that is, 
\begin{equation} 
T_j^1=1, \;\;T_j^2\neq 1,\;\; T_j^3\neq 1, \;\; j=1,2. 
\end{equation}
This condition will make the soliton $S_1$ only to be unaffected at the end of
the 
three soliton collision process. Then the equations corresponding to this
condition are
\bes
\bear
A_{1R}+A_{2R}x-A_{2I}y+A_{3R}(x^2-y^2)-2A_{3I}xy+A_{4R}x+A_{4I}y
+A_{5R}(x^2+y^2)\nonumber\\
+A_{6R}(x^3+xy^2)-A_{6I}(x^2y+y^3)+A_{7R}(x^2-y^2)\nonumber\\
+2A_{7I}xy+A_{8R}(x^3+xy^2)+A_{8I}(x^2y+y^3) 
+A_{9R}(x^2+y^2)^2 = 0, \\
A_{1I}+A_{2R}y+A_{2I}x+2A_{3R}xy+A_{3I}(x^2-y^2)+A_{4I}x-A_{4R}y
+A_{5I}(x^2+y^2)\nonumber\\
+A_{6I}(x^3+xy^2)+A_{6R}(x^2y+y^3)-2A_{7R}xy\nonumber\\
+A_{7I}(x^2-y^2)-A_{8R}(x^2y+y^3)+A_{8I}(x^3+xy^2) 
+A_{9I}(x^2+y^2)^2 = 0,\\
B_{1R}+B_{2R}x-B_{2I}y+B_{3R}(x^2-y^2)-2B_{3I}xy+B_{4R}x+B_{4I}y
+B_{5R}(x^2+y^2)\nonumber\\
+B_{6R}(x^3+xy^2)-B_{6I}(x^2y+y^3)+B_{7R}(x^2-y^2)\nonumber\\
+2B_{7I}xy+B_{8R}(x^3+xy^2)+B_{8I}(x^2y+y^3)  
+B_{9R}(x^2+y^2)^2 = 0, \\
B_{1I}+B_{2R}y+B_{2I}x+2B_{3R}xy+B_{3I}(x^2-y^2)+B_{4I}x-B_{4R}y
+B_{5I}(x^2+y^2)\nonumber\\
+B_{6I}(x^3+xy^2)+B_{6R}(x^2y+y^3)-2B_{7R}xy\nonumber\\
+B_{7I}(x^2-y^2)-B_{8R}(x^2y+y^3)+B_{8I}(x^3+xy^2) 
+B_{9I}(x^2+y^2)^2 = 0,
\eear
\ees
where we have taken $\left(\frac{\alpha_3^{(1)}}{\alpha_3^{(2)}}\right)$ $=$ $x+iy$,
the subscripts $\{lR\}$ and $\{lI\}$, $l=1,2,...,9$, represent 
the real and imaginary 
parts respectively. The expressions for $A_i$'s and $B_i$'s are lengthy but can
be obtained straightforwardly (by making use of (56) and the expressions (45a)) 
and so we do not present them here.
Solving these overdetermined system of equations for $x$ and $y$ will give the 
suitable ratio $\left(\frac{\al_3^{(1)}}{\al_3^{(2)}}\right)$  
for which the shape restoring property of one of the solitons $S_1$ 
only arises in a 
three soliton collision process.

Though we have not investigated the problem of existence of solutions of the
equations (57), one can make a numerical search and identify suitable values of
$x$ and $y$ to demonstrate the shape restoration property. For example 
in Fig. 7 with the parameters fixed at
$k_1=1+i$, $k_2 = 1.5-0.5i$, $k_3 = 2-i$, 
$\al_1^{(1)}=\al_1^{(2)}=\al_2^{(2)}=1$, $\al_2^{(1)}=\frac{39+80i}{89}$, 
$\al_3^{(1)}=1.19$, $\al_3^{(2)}=\frac{39+80i}{89}$, we have demonstrated the shape restoration property.
We find that while the amplitudes of the two of the solitons 
($S_2$ and $S_3$) change after interaction, the amplitude of the soliton 
$S_1$ remains  unchanged during the interaction process.

\begin{figure}
\caption{Shape restoring property of soliton 1 ($S_1$) during its collision with 
the other two solitons, soliton 2 ($S_2$) and soliton 3 ($S_3$), for the choice of
the parameters $k_1=1+i$, $k_2 = 1.5-0.5i$, $k_3 = 2-i$, 
$\al_1^{(1)}=\al_1^{(2)}=\al_2^{(2)}=1$, $\al_2^{(1)}=\frac{39+80i}{89}$, 
$\al_3^{(1)}=1.19$, $\al_3^{(2)}=\frac{39+80i}{89}$ .}
\end{figure}
In the above analysis we have required the complete restoration property of
soliton $S_1$. However, it is also possible to require the intensity alone to be
restored. In this case, the condition (56) can be modified as 
\bear
|T_j^1|=1,\;\;|T_j^2| \neq 1,\;\;|T_j^3| \neq 1,\;\;j=1,2,
\eear
leading to a set of two complicated equations for $x$ and $y$ (which are too
lengthy to be presented here). Solving them we
can find $x$ and $y$. Note that the quantities $x$ and $y$ correspond to 
the real and imaginary parts of the ratio of the parameters $\al_3^{(1)}$ and 
$\al_3^{(2)}$, so that for every choice of $x$ and $y$ there exists a large set
of $\al_3^{(1)}$ and $\al_3^{(2)}$ values for which shape restoration property
holds good. 
 
One might also go a step further and demand that the phase shift $\Phi^1$ or the
changes in the relative separation distances $\Delta t_{12}$ and $\Delta t_{13}$
vanish. These will give additional constraints on the choice of parameters
$\al_3^{(1)}$ and $\al_3^{(2)}$. These considerations require separate study and
we have not pursued them here. It is obvious that such shape changing and shape
restoring collision properties of the optical solitons in integrable CNLS equations, 
exhibiting a redistribution of intensity among the three solitons in the two modes, 
will have considerable technological applications both in optical communications
including wavelength division multiplexing, optical switching devices, etc., and
optical computation, for example in constructing logic gates [18, 19]. 
\subsection{Three-soliton solution of multicomponent CNLS equations and shape
changing collisions}
The above analysis on the three soliton collision in 2-CNLS equations can be
extended straightforwardly to the 3-soliton solution (14) of $N$-CNLS equations,
with arbitrary $N$, including $N=3$. One can identify that shape changing
collision occurs here also but with lot more possibilities for redistribution of
intensities in contrast to the 2-CNLS case. The quantities characterizing the
collision process here also are the intensity redistribution, amplitude
dependent phase shifts and relative separation distances between the solitons, as
explained in the 2-CNLS case.

We also note that as the number of components increases from two to some
arbitrary $N$ ($N > 2$), the different possibilities for redistribution of 
intensity among them also increase in a manifold way.  The corresponding transition matrix, measuring this
redistribution, is found to be similar to Eqs. (45) with the redefinition of 
$\kappa_{il}$'s as given in Eq. (14b) along with the index $j$ running from
1 to $N$ instead of 1 to 2. The other factors, amplitude dependent phase shifts
and change in relative separation distances, also bear the same form given by
Eqs. (46) and (47), respectively, with this redefinition. 

As to the shape restoration property one has to again solve the equations 
\bear
T_j^1=1, \;\; T_j^2 \neq 0,\;\;T_j^3 \neq 0, \;\;j=1,2,3,...,N.
\eear
Alternatively for intensity restoration the conditions are 
\bear
|T_j^1|=1, \;\; |T_j^2| \neq 0,\;\;|T_j^3| \neq 0, \;\;j=1,2,3,...,N.
\eear

Extending the above analysis, it is clear that, carrying out an asymptotic
analysis of 4-soliton solution given in Appendix A, it is possible to restore
the shape of two of the solitons at the maximum, which can be further
generalized to arbitrary $N$ soliton case, in which it is possible to restore the
shape of $N-2$ of the solitons. We have checked in this case also from the
asymptotic analysis the soliton interaction is pair-wise and we conjecture that
this should be true for the arbitrary $N$-soliton case as well.
\section{Multisoliton solutions as logic gates}
The state vectors and LFTs introduced in Sec. VI and the shape changing
pair-wise collision nature of bright solitons mentioned in Sec. VII can be
profitably used to look at the multisoliton solutions of CNLS equations as
various logic gates. We believe that such an approach provides an alternative
point of view of shape changing soliton collisions to construct logic gates as
discussed in Ref. [19]. The present point of view may have its own advantage as
system initial conditions are chosen suitably to generate specific forms of multisolitons to
represent logic gates may be much easier from a practical point of view,
including replication, compared to constructing them through predetermined
independent soliton collisions. In the following we will demonstrate this idea
for the case of the 2-CNLS as an example.
\subsection{Three soliton solution and state restoration property }
The shape restoration of a particular soliton in arbitrary state associated with
the three soliton solution has been
discussed in Sec. VII F. Particularly this can be well appreciated with respect
to binary logic states. For example, if we consider the soliton $S_1$ is in 
1 state with the state value $\rho_{1,2}^{1-} = 1$, it implies
\bear
\frac{\al_1^{(1)}}{\al_1^{(2)}} =1.
\eear
To obtain this we choose, $\al_1^{(1)}$ $=$ $\al_1^{(2)}$ $=$ $1$.
For simplicity we require $S_2$ to be in 0 state before interaction. From
the asymptotic expressions (39), this can be achieved by choosing the ratio  
$\frac{\al_2^{(1)}}{\al_2^{(2)}}$  as 
\bear
\frac{\al_2^{(1)}}{\al_2^{(2)}}=\frac{k_1+k_1^*}{2k_2+k_1^*-k_1}.
\eear
Now in order to restore the state of $S_1$ after two collisions, 
we have to allow the outcome of $S_1$ resulting 
after the first collision, which may be called soliton $S_1'$, 
to interact with soliton $S_3$ having 
a state inverse to the above 0
state. This state for $S_3$ can be identified from its asymptotic form before
interaction given in Eq. (40). The resulting condition can be shown to be
\bes
\bear
\frac{\al_3^{(1)}}{\al_3^{(2)}}&=&\frac{n}{d},\\
n&=&
-(\al_2^{(1)}+\al_2^{(2)})\al_2^{(2)*}A+\kappa_{22}(k_1-k_1^*-2k_3)B
+2\al_2^{(2)}\al_2^{(2)*}C\nonumber\\
&&-\al_2^{(2)}(\al_2^{(1)*}+\al_2^{(2)*})D
+|\al_2^{(1)}+\al_2^{(2)}|^2E,\\
d &=&
(\al_2^{(1)}+\al_2^{(2)})\al_2^{(1)*}A-\kappa_{22}(k_1+k_1^*)B
-2\al_2^{(2)}\al_2^{(1)*}C\nonumber\\
&&-(\al_2^{(1)*}+\al_2^{(2)*})\al_2^{(2)}D,\\
\mbox{where}\qquad \qquad \qquad\nonumber\\
A&=& (k_1+k_1^*)(k_3+k_1^*)(k_1+k_2^*),\\
B&=& (k_2+k_1^*)(k_3+k_2^*)(k_1+k_2^*),\\
C&=& (k_2+k_1^*)(k_3+k_1^*)(k_1+k_2^*),\\
D &=& (k_2+k_1^*)(k_3+k_2^*)(k_1+k_1^*),\\
E &=& (k_3+k_2^*)(k_1+k_1^*)(k_3+k_1^*).
\eear
\ees
In the above equation choosing the parameters satisfying conditions (61) and
(62) one can fix $\left(\frac{\al_3^{(1)}}{\al_3^{(2)}}\right)$ 
suitably in order to restore the state of soliton $S_1$. 
Thus the three soliton solution given by Eq. (10) having the specific choice of 
parameters  specified by Eqs. (61)-(63) corresponds to 
the state restoration of soliton $S_1$.
\subsection{Four-soliton solution and COPY gate}
Extending the above procedure, we can now
consider the four-soliton solution given in Appendix A, and identify it as (i) a
COPY gate or (ii) a ONE gate or (iii) a NOT gate studied in Ref. [19] for
suitable choices of the arbitrary parameters. As
an example, let us consider copying 1 state of $S_1$ to the output state of 
soliton $S_4$. This
requires the following steps.
\begin{enumerate}
\item
We consider the four soliton collision process in which the soliton $S_1$
collides with the soliton $S_2$ first and then with the soliton $S_3$ and finally
with the soliton $S_4$. This sequence of collision follows from the condition
$k_{1I}>k_{2I}>k_{3I}>k_{4I}$.
\item
Consider for convinience $S_1$ to be in the 0 state, the so-called actuator state
[19]. This requires $\frac{\al_1^{(1)}}{\al_1^{(2)}}=0$, which can be obtained
by choosing $\al_1^{(1)}=0$ and $\al_1^{(2)}$ as arbitrary.
\item
Assign 1 state to soliton $S_2$ before interaction, for which we need 
\bear
\frac{\al_2^{(1)}}{\al_2^{(2)}}=\frac{k_2-k_1}{k_2+k_1^*}.
\eear
\item
After its collision with $S_2$ as a result of shape changing collision the outcoming
state of $S_1$ (say $S_1'$) will be altered.
\item
Now let us allow the third soliton in the four soliton solution to interact with
$S_1'$ which changes the state $S_1'$  to $S_1''$.
\item
Finally $S_4$ is allowed to interact with $S_1''$. From the asymptotic analysis,
we identify the state of soliton $S_4$ after interaction as 
\bear
\rho_{1,2}^{4+}=\frac{\al_4^{(1)}}{\al_4^{(2)}}.
\eear
We impose the condition on this state that this should be in the state of $S_2$
before interaction. Thus the  parameters $\al_4^{(1)}$ and $\al_4^{(2)}$ 
of soliton $S_4$ get
fixed depending upon the input state of $S_2$.
\item
The asymptotic analysis of the four soliton solution given in Appendix A
results in following condition for $S_4$ to be in one state after
interaction.
\bear
\frac{T_1^4\;A_1^{4-}}{T_2^4\;A_2^{4-}}=1,
\eear
where $T_1^4$ and $T_2^4$ are the transition elements of $S_4$ in the modes $q_1$
and $q_2$ respectively. Here $A_1^{4-}k_{4R}$ and $A_2^{4-}k_{4R}$ are the amplitudes of
soliton $S_4$ before interaction in the two modes respectively.
\item
If we flip the input state of $S_2$ from 1 to 0 state by suitably choosing the
$\rho_{1,2}^{2-}$'s parameters then the condition on soliton $S_4$'s output will become
\bear
\frac{T_1^4\;A_1^{4-}}{T_2^4\;A_2^{4-}}=0.
\eear
\item In the above two Eqs. (66) and (67) only free parameters are $\al_3^{(1)}$
and $\al_3^{(2)}$. In principle, we can solve these two complex equations to
obtain the free complex parameters $\al_3^{(1)}$ and $\al_3^{(2)}$. Then for the
given choice of parameters the state of the incoming soliton $S_2$ can be copied
on to the outgoing soliton $S_4$.
\end{enumerate}
Thus a four soliton collision process with the above premise is equivalent to a
COPY gate. 
Similar procedure can be extended to other gates mentioned above as well. One
can extend this idea further to identify a FANOUT gate from a 5-soliton
solution. It appears that one can pursue the idea ultimately to identify the
NAND gate itself as a multisoliton solution following the construction of
Steiglitz in Ref. [19]. Fuller details will be reported elsewhere. 
\section{Bright soliton solutions and partially coherent solitons}
As mentioned in the introduction, the recent observations by several authors
[11, 12, 25]
have shown  that 
the $N$-CNLS equations (1) can support $N$-PCSs solutions. In general,
these PCSs are said to be special cases of the so-called 
multisoliton complexes [2] which are nonlinear
superposition of fundamental bright solitons. It has also been
demonstrated that these PCSs are formed only if the number of components in Eq.
(1) is equal to the
number of solitons. Then it is quite natural to look for the 2-PCS, 3-PCS, 4-PCS,
etc., as special cases of the two-soliton solution of the 2-CNLS, three-soliton
solution of the 3-CNLS, four- soliton solution of the 4-CNLS equations etc.,
respectively, deduced in Secs. III and IV. In the following, we 
indeed show that the  PCSs reported in Refs.[11, 12, 25] 
result as special cases, that is specific choices of
some of the arbitrary complex parameters,
from the bright soliton solutions of CNLS equations discussed in Secs. III and
IV and thereby showing the origin of the various interesting properties of the
PCS solutions.
\subsection{ 2-PCS : A special case of the bright two-soliton solution of 2-CNLS equations}
Let us consider stationary limit of the two-soliton solution of the 2-CNLS equations 
(Manakov system) given by Eq. (8),  that is, 
 $k_{nI}=0$, for the special choice of the parameters, 
$\al_1^{(1)}=e^{\eta_{10}}$, $\al_2^{(2)}=-e^{\eta_{20}}$,
$\al_1^{(2)}=-\al_2^{(1)}=0$, where $\eta_{j0}$'s, $j=1,2$, are now restricted as
real constants.
Then Eq. (8) becomes
\begin{subequations}
\bear
q_1 & = & \left(e^{\eta_1}+\frac{\mu(k_{1R}-k_{2R})e^{\eta_1+\eta_2+
\eta_2^*}}{4k_{2R}^2(k_{1R}+k_{2R})}\right)/\tilde{D},\\
q_2 & = & \left(-e^{\eta_2}+\frac{\mu(k_{1R}-k_{2R})e^{\eta_1+\eta_1^*+
\eta_2}}{4k_{1R}^2(k_{1R}+k_{2R})}\right)/\tilde{D},
\eear
where
\bear
 \tilde{D}&=& 1+\mu\left[\frac{e^{\eta_1+\eta_1^*}}{4k_{1R}^2}+
\frac{e^{\eta_2+\eta_2^*}}{4k_{2R}^2}\right]+\frac{\mu^2(k_{1R}-k_{2R})^2
e^{\eta_1+\eta_1^*+\eta_2+\eta_2^*}}
{16k_{1R}^2k_{2R}^2(k_{1R}+k_{2R})^2},
\eear
and
\bear
\eta_j=k_{jR}(t+ik_{jR}z)+\eta_{j0},\;\;j=1,2.
\eear
\end{subequations}
This stationary solution can be easily  identified as the 2-PCS 
expression (13-15) given
in Ref. [12] with the identification of $\bar{x}_j$'s  as
$\bar{t}_{j}$'s, $j=1,2$, 
\begin{subequations} 
\begin{eqnarray}
\bar{t_1}&=&t-t_1=t+\frac{\eta_{10}}{k_{1R}}+\frac{1}{2k_{1R}}
\mbox{log}\left[\frac{\mu(k_{1R}-k_{2R})}{4k_{1R}^2(k_{1R}+k_{2R})}
\right],\\
\bar{t_2}&=&t-t_2=t+\frac{\eta_{20}}{k_{2R}}+\frac{1}{2k_{2R}}
\mbox{log}\left[\frac{\mu(k_{1R}-k_{2R})}{4k_{2R}^2(k_{1R}+k_{2R})}
\right].
\end{eqnarray}
\end{subequations}

As the 2-PCS is a special case of the bright 
two-soliton solution of 2-CNLS equations, 
it is also characterized
by $\al_i^{(j)}$'s (through $\eta_{j0}$'s) and $k_{iR}$'s resulting in 
amplitude dependent phases and hence amplitude dependent relative separation distances. To be
specific, in the PCSs the change in the relative separation distance
play a predominant role in determining their shape as pointed out in [11,12]. 
These PCSs can be
classified into two types as symmetric and asymmetric depending on the relative separation
distances. Defining the relative separation distance $t_{12}$ = $t_2 - t_1$, one
can check that, for $t_{12}=0$, the PCS bears a symmetric form with respect to
its propagation direction and is known as symmetric PCS [11]. It takes an 
asymmetric
form for  $t_{12}\neq 0$ and is known as asymmetric PCS [11]. From Eqs. (69), 
the relative
separation distances for the stationary 2-PCS can be obtained as 
\bear
t_{12} = t_2-t_1 = \frac{\eta_{10}}{k_{1R}}-
\frac{\eta_{20}}{k_{2R}}+\frac{1}{2k_{1R}}
\mbox{log}\left[\frac{\mu(k_{1R}-k_{2R})}{4k_{1R}^2(k_{1R}+k_{2R})}\right]
-\frac{1}{2k_{2R}}
\mbox{log}\left[\frac{\mu(k_{1R}-k_{2R})}{4k_{2R}^2(k_{1R}+k_{2R})}\right].\nonumber\\
\eear
Typical forms of symmetric and asymmetric stationary 2-PCS are shown in Fig. 8,
which in similar to the ones in Ref. [12].
\begin{figure}
\caption{
Typical 2-PCS forms for the Manakov system for $z=0$ see Eqs. (68),
with $k_1=1.0$ and $k_2=2.0$:
(a) Symmetric case ($t_{12}=0$), (b) Asymmetric case ($t_{12}=1$).}
\end{figure}
\subsection{ 3-PCS: A special case of the bright
three-soliton solution of 3-CNLS equations}
Since it has been observed that the PCS solutions exist when the 
number of components
is equal to the number of solitons propagating in the system, we consider next the
3-soliton solution of the 3-CNLS equations 
in order to show that the  3-PCS as
a special case of the 3-soliton solution. Thus considering the stationary limit 
$k_{nI} = 0$, $n=1,2,3,$ of the 3-soliton solution of 3-CNLS equations given by
Eq. (14) with $N=3$, and making the following parametric choice
\bear
\alpha_1^{(1)}=e^{\eta_{10}}, \alpha_2^{(2)}=-e^{\eta_{20}},  
\alpha_3^{(3)}=e^{\eta_{30}},  
\alpha_1^{(2)}=\alpha_1^{(3)}=\alpha_2^{(1)}=\alpha_2^{(3)}=
\alpha_3^{(1)}=\alpha_3^{(2)}=0,
\eear
where $\eta_{j0}$'s, $j=1,2,3$, are restricted to  real parameters,
we obtain 
\begin{subequations}
\bear
q_1 & = & \left[e^{\eta_1}+\frac{\mu(k_{1R}-k_{2R})e^{\eta_1+\eta_2+
\eta_2^*}}{4k_{2R}^2(k_{1R}+k_{2R})}+\frac{\mu(k_{1R}-k_{3R})e^{\eta_1+\eta_3+
\eta_3^*}}{4k_{3R}^2(k_{1R}+k_{3R})} \right.\nonumber\\
& &\qquad \left.+\frac{\mu^2(k_{2R}-k_{1R})(k_{3R}-k_{1R})(k_{3R}-k_{2R})^2
e^{\eta_3+\eta_3^*+\eta_2+\eta_2^*+\eta_1}}{16k_{2R}^2k_{3R}^2
(k_{2R}+k_{1R})(k_{3R}+k_{1R})(k_{3R}+k_{2R})^2}\right]/\tilde{D_1},\\
q_2 & = & \left[-e^{\eta_2}+\frac{\mu(k_{1R}-k_{2R})e^{\eta_1+\eta_1^*+
\eta_2}}{4k_{1R}^2(k_{1R}+k_{2R})}+\frac{\mu(k_{3R}-k_{2R})e^{\eta_3+\eta_3^*+
\eta_2}}{4k_{3R}^2(k_{3R}+k_{2R})}\right.\nonumber\\
& &\qquad\left.+\frac{\mu^2(k_{2R}-k_{1R})(k_{3R}-k_{2R})(k_{3R}-k_{1R})^2
e^{\eta_3+\eta_3^*+\eta_1+\eta_1^*+\eta_2}}{16k_{1R}^2k_{3R}^2
(k_{2R}+k_{1R})(k_{3R}+k_{2R})(k_{3R}+k_{1R})^2}\right]/\tilde{D_1},\\
q_3 & = &\left[e^{\eta_3}+\frac{\mu(k_{3R}-k_{1R})e^{\eta_1+\eta_1^*+
\eta_3}}{4k_{1R}^2(k_{1R}+k_{3R})}+\frac{\mu(k_{3R}-k_{2R})e^{\eta_2+\eta_2^*+
\eta_3}}{4k_{2R}^2(k_{3R}+k_{2R})}\right.\nonumber\\
& &\qquad\left.+\frac{\mu^2(k_{3R}-k_{1R})(k_{3R}-k_{2R})(k_{2R}-k_{1R})^2
e^{\eta_2+\eta_2^*+\eta_1+\eta_1^*+\eta_3}}{16k_{1R}^2k_{2R}^2
(k_{3R}+k_{1R})(k_{3R}+k_{2R})(k_{2R}+k_{1R})^2}\right]/\tilde{D_1}.
\eear
Here, 
\bear
\tilde{D_1}&=& 1+\mu\left[\frac{e^{\eta_1+\eta_1^*}}{4k_{1R}^2}+
\frac{e^{\eta_2+\eta_2^*}}{4k_{2R}^2}+\frac{e^{\eta_3+\eta_3^*}}
{4k_{3R}^2}\right] 
+\frac{\mu^2(k_{1R}-k_{2R})^2e^{\eta_1+\eta_1^*+\eta_2+\eta_2^*}}
{16k_{1R}^2k_{2R}^2(k_{1R}+k_{2R})^2}\nonumber\\
&&+\frac{\mu^2(k_{1R}-k_{3R})^2e^{\eta_1+\eta_1^*+\eta_3+\eta_3^*}}
{16k_{1R}^2k_{3R}^2(k_{1R}+k_{3R})^2}
+
\frac{\mu^2(k_{3R}-k_{2R})^2e^{\eta_2+\eta_2^*+\eta_3+\eta_3^*}}
{16k_{2R}^2k_{3R}^2(k_{2R}+k_{3R})^2}\nonumber\\
&&+\left[\frac{\mu^3(k_{2R}-k_{1R})^2(k_{3R}-k_{1R})^2(k_{3R}-k_{2R})^2
e^{\eta_1+\eta_1^*+\eta_2+\eta_2^*+\eta_3+\eta_3^*}}
{64 k_{1R}^2k_{2R}^2k_{3R}^2(k_{1R}+k_{2R})^2(k_{1R}+k_{3R})^2(k_{2R}+k_{3R})^2}\right].
\nonumber\\
\eear
\end{subequations}
The above solution can be easily rewritten as Eq. (16-18) for the 3-PCS case  
given in Ref. [12]. As in the case of 2-PCS, here also we identify $\bar{x}_j$'s
given in Ref. [12] as $\bar{t}_j$'s, $j=1,2,3$, which are defined below:
\bes
\begin{eqnarray}
\bar{t_1} &=& t-t_1=t+\frac{\eta_{10}}{k_{1R}}+\frac{1}{2k_{1R}}\mbox{log} 
\left[\frac{\mu(k_{2R}-k_{1R})(k_{3R}-k_{1R})}{4k_{1R}^2(k_{1R}+k_{2R})
(k_{1R}+k_{3R})}\right],\\
\bar{t_2} &=& t-t_2=t+\frac{\eta_{20}}{k_{2R}}+\frac{1}{2k_{2R}}\mbox{log} 
\left[\frac{\mu(k_{2R}-k_{1R})(k_{3R}-k_{2R})}{4k_{2R}^2(k_{1R}+k_{2R})
(k_{2R}+k_{3R})}\right],\\
\bar{t_3} &=& t-t_3=t+\frac{\eta_{30}}{k_{3R}}+\frac{1}{2k_{3R}}\mbox{log} 
\left[\frac{\mu(k_{3R}-k_{1R})(k_{3R}-k_{2R})}{4k_{3R}^2(k_{1R}+k_{3R})
(k_{2R}+k_{3R})}\right].
\eear
\end{subequations}
These 3-PCSs can also be classified as symmetric and asymmetric as in the case
of 2-PCSs. The stationary
3-PCS is symmetric when $t_{12} =  t_{13}=0$ and asymmetric
otherwise. In Fig.9 we have shown the symmetric and asymmetric 
3-PCS solutions.
\begin{figure}[h]
\caption{Typical 3-PCS forms for the integrable 3-CNLS system for $z=0$ with
$k_1=1.0$, $k_2=0.5$ and $k_3=0.2$, see 
Eqs. (72):
(a) Symmetric case ($t_{12}= t_{13}=0$), (b) Asymmetric case ($t_{12}=1,\; 
t_{13}=2$).}
\end{figure}
\subsection{4-PCS: A special case of the 
four-soliton solution of 4-CNLS equations}

In a similar fashion as in the above two cases, 
the 4-soliton solution of the 4-CNLS equations given in
Appendix A, with $N=4$ can also be shown to reduce to 4-PCS given by Eq. (19-23) in 
Ref. [12] 
by choosing,  $k_{nI}=0$,
$\al_1^{(1)}=e^{\eta_{10}}$, $\al_2^{(2)}=-e^{\eta_{20}}$, 
$\al_3^{(3)}=e^{\eta_{30}}$, $\al_4^{(4)}=-e^{\eta_{40}}$, 
$\al_i^{(j)}=0$, $i, j=1,2,3,4$, $i \neq j$. Since it is straightforward but
lengthy to
write down the form, we desist from presenting the solution here. Here the 
$t_j$'s, $j=1,2,3,4$, are
defined as
\bes
\bear
t_1 &=& -\frac{\eta_{10}}{k_{1R}}-\frac{1}{2k_{1R}}\mbox{log}\left[
\frac{\mu(k_{2R}-k_{1R})(k_{3R}-k_{1R})(k_{4R}-k_{1R})}
{4k_{1R}^2(k_{2R}+k_{1R})(k_{3R}+k_{1R})(k_{4R}+k_{1R})} \right], \\
t_2 &=& -\frac{\eta_{20}}{k_{2R}}-\frac{1}{2k_{2R}}\mbox{log}\left[
\frac{\mu(k_{2R}-k_{1R})(k_{3R}-k_{2R})(k_{4R}-k_{2R})}
{4k_{2R}^2(k_{2R}+k_{1R})(k_{3R}+k_{2R})(k_{4R}+k_{2R})} \right], \\
t_3 &=& -\frac{\eta_{30}}{k_{3R}}-\frac{1}{2k_{3R}}\mbox{log}\left[
\frac{\mu(k_{3R}-k_{1R})(k_{3R}-k_{2R})(k_{4R}-k_{3R})}
{4k_{3R}^2(k_{3R}+k_{1R})(k_{3R}+k_{2R})(k_{4R}+k_{3R})} \right], \\
t_4 &=& -\frac{\eta_{40}}{k_{4R}}-\frac{1}{2k_{4R}}\mbox{log}\left[
\frac{\mu(k_{4R}-k_{1R})(k_{4R}-k_{2R})(k_{4R}-k_{3R})}
{4k_{4R}^2(k_{4R}+k_{1R})(k_{4R}+k_{2R})(k_{4R}+k_{3R})} \right]. 
\eear
\ees
Here also the symmetric PCS results for 
$t_{ij} = 0$, $j>i$, $i,j=1,2,3,4$,  and asymmetric PCS for 
$ t_{ij}$ $\neq 0$, $j>i$. 

Extending this
idea to arbitrary $N$, it is clear that the $N$-PCS is a special case of the 
$N$-soliton
solution of $N$-CNLS equations (1). It has been noticed in Refs. [11, 12] that these
PCSs are of variable shape. The reason for the variable shape can be traced
naturally to the nontrivial dependence of phases on the complex parameters
$\al_i^{(j)}$'s as shown above. 
Thus it is clear that any change in the amplitude will affect the phase part of
the solitons and vice-versa. Since we have explicitly shown that $N$-PCSs are
special cases of bright N-soliton solutions of $N$-CNLS equations, 
they possess variable shape
as a consequence of the shape dependence on the $\al_i^{(j)}$ parameters.
\subsection{Propagation of partially coherent solitons and their collision
properties}
The intriguing collision properties of the partially coherent solitons reported
in Refs. [11, 12] can be
well understood by writing down the expression for PCSs with nonvanishing
$k_{nI}$'s, that is nonstationary special cases of multicomponent 
higher order bright soliton
solutions discussed in Secs. III and IV.  For the nonstationary PCSs we can choose as a
special case 
the complex parameters $\al_i^{(j)}$'s ($i \neq j$) to be
functions of $k_{nI}$'s such that they vanish as $k_{nI} = 0$. 
As we make these $k_{nI}
\neq 0$, then $\al_i^{(j)}$'s ($i \neq j$) also vary, thereby making the
collision scenario interesting. We can consider both the cases of equal and
unequal velocities, which exhibit similar behaviours.

As a first example, we consider the propagation of the  2-PCS comprising two
solitons with equal velocities ($k_{1I}=k_{2I}$) in PR media. Its
propagation can be studied by choosing (for illustrative purpose) 
$\alpha_1^{(2)}=k_{1I}$ and $\alpha_2^{(1)}=(0.25+1.02i)k_{2I}$ as
functions of velocities ($k_{jI},j=1,2$) such that they vanish when $k_{jI} =
0$, $j=1,2$. This is shown in Fig. 10 for the parameters
$\alpha_1^{(1)}=2.0+i$,$\;$
$\;$$\alpha_2^{(2)}=1$,$\;$
$k_1=1.0+i$, and $k_2=2.0+i$. 
For unequal velocity case ($k_{1I}\neq k_{2I}$), the PCS collision
is shown in Fig. 11 for the parametric choice 
$\alpha_1^{(1)}=1.0$,$\;$$\alpha_1^{(2)}=k_{1I}$,$\;$
$\alpha_2^{(1)}=-(\frac{22+80i}{89})k_{2I}$,$\;$$\alpha_2^{(2)}=-2.0$,$\;$
$k_1=1.0+i$, and $k_2=2.0-i$. This can also be viewed as collision of two 1-PCS
which are spread among the two components, which are travelling with equal but
opposite velocities.
\begin{figure}[h]
\caption{Intensity profiles showing the collision scenario of two 1-PCSs, with
equal velocities,
at (a) $z=-5$ and (b) $z=5$, given by special choice of parameters (as given in
text) in the 2-soliton solution of the Manakov system.}
\end{figure} 
\begin{figure}[h]
\caption{Intensity profiles showing the collision scenario of two 1-PCSs,
moving with equal but opposite velocities, at (a) $z=-5$ and (b) $z=5$, given by special choice of parameters (as given in
text) in the 2-soliton solution of the Manakov system.}
\end{figure}

Now let us consider the collision of 2-PCS and 1-PCS in PR media. This is
equivalent to the three soliton collision in the 3-CNLS system with specific choice
of parameters. We consider the case, 
in
which the complex parameters 
$\alpha_1^{(2)}$, $\alpha_1^{(3)}$ , $\alpha_2^{(1)}$ , $\alpha_2^{(3)}$,
$\alpha_3^{(1)}$, $\alpha_3^{(2)}$ are nonvanishing and as functions of
$k_{nI}$'s, $n=1,2,3$. Then the resulting asymptotic forms of the 
3-PCS propagation is shown in Fig. 12 for the
parametric choice 
$\alpha_1^{(1)}=1.0 $,$\;$
$\alpha_1^{(2)}=\alpha_1^{(3)}=k_{1I}$,$\;$
$\alpha_2^{(1)}=-0.5 k_{2I}$,$\;$ 
$\alpha_2^{(2)}=0.25$,$\;$
$\alpha_2^{(3)}=0.02 k_{2I}$,$\;$ 
$\alpha_3^{(1)}=-(\frac{22+80i}{89})k_{3I}$, $\;$
$\alpha_3^{(2)}=2 k_{3I}$,$\;$
$\alpha_3^{(3)}=-2 $,$\;$ 
$k_1=1.0+i$,$\;$  $k_2=1.5-i$, $\;$and $k_3=2.0-i$ . 
\begin{figure}[h]
\caption{Intensity profiles showing the collision scenario of 2-PCS with 1-PCS
at (a) $z=-4$ and (b) $z=4$ given by special choice of parameters (as given in
text) in the 3-soliton solution of Eq.(1) with $N=3$.}
\end{figure} 
In the above
figures it can be verified that the total intensity of the individual solitons
comprising the PCS is conserved. 

The above analysis on PCS propagation clearly shows that, there will be variation
in the shape of the PCS during its collision with other PCSs. The explanation for
this result follows from the shape changing (intensity redistribution) nature of
fundamental bright soliton collision of the
integrable CNLS equations, explained in Section V. Further, we have also
observed that the collision of two PCSs each comprising $m$ and $n $ soliton
complexes, respectively, such that $m+n=N$ studied in Refs. [11, 12, 25], is equivalent to the interaction of
$N$ fundamental bright solitons (for suitable specific choice of parameters) represented by
the special case of $N$-soliton solution of the $N$-CNLS system. It should also be
noted that in the collision process the total intensity of individual solitons
comprising the $N$-PCS is conserved. This is due to the complete integrable nature
of the $N$-CNLS equations (1). 
\subsection{Multisoliton Complexes}
In the above we have considered the CNLS equations with number of components
(say p) is equal to the number of fundamental solitons (say q). This is only a
special case of the multisoliton complexes  and has been much discusssed
recently. However, the results are scarce for the case $p \neq q$, except for
the work of Sukhorukov and Akhmediev [25], where the incoherent soliton
collision is demonstrated numerically. To elucidate the
understanding we present a form of the three soliton complex in which three
solitons are spread among the two components, by suitably choosing the
parameters in the explicit expression, Eq. (10). This has been shown in Fig. 13
with the parameters chosen as 
$\alpha_1^{(1)}=\alpha_1^{(2)}=1.0$,$\;$
$\alpha_2^{(1)}=0.5$, $\alpha_2^{(2)}=0.25$, 
$\alpha_3^{(1)}=\frac{22+80i}{89}$, $\alpha_3^{(2)}=-2$,
$k_1=1.0+i$, and $k_2=1.5+i$. 
\begin{figure}[h]
\caption{Intensity profiles of a multisoliton complex comprising three solitons
spread up in two components propagating in photorefractive media: A special case
of the 
three soliton solution (10) of the integrable 2-CNLS system for the parameters
chosen as in the text, (a) at $z=-5$ and (b) at $z=5$.}
\end{figure}
From the figure and the analysis of
the soliton interaction it is clear that the shape of these complexes strongly
dependent on the $\al_i^{(j)}$'s along with $k_j$'s which determine how the solitons
are spread up among the components.
For the same case there exists various forms of multisoliton complexes
depending on the spreading up of solitons in the two components. As a
consequence of this multisoliton complexes will possess a rich variety of
structures in comparison with the PCSs.

\section{Conclusion}
We conclude this paper  by stating that the
collision processes of solitons in coupled nonlinear Schr{\"o}dinger equations
lead to very many exciting novel properties and potential applications. The
novel properties include shape changing intensity redistributions, amplitude
dependent phase shifts and relative separation distances, within the pair-wise
collision mechanism of solitons. Interestingly, it is identified that the
intensity redistribution characterising the shape changing collision process in
$N$-CNLS equations can be written as a generalized linear fraction
transformation. This will giver further impetus in constructing multistate
logic, multi-input logic
gates, memory storage devices and so on by using soliton interactions. The
implication of these properties require further deep investigations. Further,
viewing the recently much discussed objects multisoliton complexes, partially
coherent solitons as special cases of the bright soliton solution enhances the
understanding of their various properties. We expect the interaction study
presented here will shine more light on spatial soliton propagation in (1+1)D
 photorefractive planar waveguides.

\section*{Acknowledgment}
\noindent The work reported here has been supported by the Department of Science
and Technology, Govt. of India and the Council of Scientific and Industrial 
Research (CSIR), Government of India in the form of research projects to M.L. 
T.K. wishes to acknowledge CSIR for the award of a Senior Research Fellowship.

\appendix 
\section{Four-soliton solution}
In this appendix for completeness, we present the form of the four-soliton
solution of the 2-CNLS equations by generalizing the two- and three- soliton 
solutions of it, which can be obtained by terminating the power series as  
\bes
\bear
g^{(j)} &=& \chi g_1^{(j)}+ \chi^3 g_3^{(j)} +\chi^5 g_5^{(j)} 
+\chi^7 g_7^{(j)},\\
f &=& 1+ \chi^2 f_2+ \chi^4 f_4+ \chi^6 f_6 +\chi^8 f_8,
\eear
\ees
and solving the resulting set of linear partial 
differential equations recursively. It 
can be written as

\bes
\bear
q_s=\frac{N^{(s)}}{D},\;\;s=1,2,
\eear
where
\bear
N^{(s)} &=& \sum_{i=1}^4 \al_i^{(s)}e^{\eta_i}+
\left(\frac{1}{2}\right)
\sum_{\vspace*{-5cm}\begin{array}{c} \vspace{-10pt}
       i,j,l=1 \\ \vspace{-10pt}
      (i\ne l)  \end{array} }^{4}
\frac{(k_l-k_i)(\al_l^{(s)}\kappa_{ij}-\al_i^{(s)}
\kappa_{lj})}{(k_j^*+k_i)(k_j^*+k_l)}e^{\eta_i+\eta_j^*+\eta_l}\nonumber\\
&&+\left(\frac{1}{12}\right)
\sum_{\vspace*{-5cm}\begin{array}{c} \vspace{-10pt}
       i,j,l,\\ \vspace{-10pt}
       m,n=1 \\  \vspace{-10pt}
      (i\ne l \ne n; \\
       j \ne m ) \end{array} }^{4} \vspace{-20 pt}
       \begin{array}{c} 
      \left[\frac{(k_n-k_i)(k_l-k_i)(k_l-k_n)(k_m^*-k_j^*)
      }{(k_j^*+k_i)(k_j^*+k_l)(k_j^*+k_n)(k_m^*+k_i)(k_m^*+k_l)(k_m^*+k_n)}
\right]{\mathbf.}\left\{ \al_i^{(s)}\left[\kappa_{lm}\kappa_{nj}-\kappa_{lj}\kappa_{nm}\right ] \right. \\
\left. +\al_n^{(s)}\left[\kappa_{lj}\kappa_{im}-\kappa_{ij}\kappa_{lm}\right]  
 +\al_l^{(s)}\left[\kappa_{nm}\kappa_{ij}-\kappa_{im}\kappa_{nj}\right] \right\}
 e^{\eta_i+\eta_j^*+\eta_l+\eta_m^*+\eta_n}
\end{array}\nonumber\\
&&-\left(\frac{1}{144}\right)\sum_{\vspace*{-5cm}\begin{array}{c} \vspace{-10pt}
       i,j,l,m,\\ \vspace{-10pt}
       n,o,p=1 \\  \vspace{-10pt}
      (i\ne l \ne n \ne p;\\ \vspace{-10pt}
      j \ne m \ne o)  \end{array} }^{4}
      \begin{array}{c}
 \frac{1}{D_1}\left[(k_p-k_i)(k_p-k_l)(k_p-k_n)(k_n-k_l)(k_n-k_i)(k_l-k_i) \right.\\
 \vspace{-10pt}
\left.(k_o^*-k_m^*)(k_o^*-k_j^*)(k_m^*-k_j^*)
 \right] \qquad \qquad \qquad \qquad 
\end{array}
\nonumber\\
&&\times\left|
\begin{array}{cccc}
\al_i^{(s)} & \al_l^{(s)} & \al_n^{(s)} & \al_p^{(s)} \\
\kappa_{ij} & \kappa_{lj} & \kappa_{nj} & \kappa_{pj} \\
\kappa_{im} & \kappa_{lm} & \kappa_{nm} & \kappa_{pm} \\
\kappa_{io} & \kappa_{lo} & \kappa_{no} & \kappa_{po}
\end{array}
\right|
e^{\eta_i+\eta_j^*+\eta_l+\eta_m^*+\eta_n+\eta_o^*+\eta_p},
\eear
where
\bear
\eta_i &=& k_i(t+i k_i z),\;\;i=1,2,3,4, \\
D_1 &=&
(k_j^*+k_i)(k_j^*+k_l)(k_j^*+k_n)(k_j^*+k_p)(k_m^*+k_i)(k_m^*+k_l)\nonumber\\
&&(k_m^*+k_n)
(k_m^*+k_p)(k_o^*+k_i)(k_o^*+k_l)(k_o^*+k_n)(k_o^*+k_p)
\eear
and
\bear
D&=& 1+\sum_{i,j=1}^4\frac{\kappa_{ij}}{k_i+k_j^*}e^{\eta_i+\eta_j^*}
+\left(\frac{1}{4}\right)\sum_{\vspace*{-5cm}\begin{array}{c} \vspace{-10pt}
       i,j,
       l,m=1 \\  \vspace{-10pt}
      (i\ne l;j \ne m )\end{array} }^{4}
\frac{(k_l-k_i)(k_m^*-k_j^*)(\kappa_{ij}\kappa_{lm}-\kappa_{im}\kappa_{lj})}
{(k_j^*+k_i)(k_j^*+k_l)(k_m^*+k_i)(k_m^*+k_l)}
e^{\eta_i+\eta_j^*+\eta_l+\eta_m^*}\nonumber\\ \vspace{50pt}
&&+\left(\frac{1}{36}\right)\sum_{\vspace*{-5cm}\begin{array}{c} \vspace{-10pt}
       i,j,l,\\ \vspace{-10pt}
       m,n,o=1 \\  \vspace{-10pt}
      (i\ne l \ne n;\\ \vspace{-10pt}
      j \ne m \ne o)  \end{array} }^{4}
\frac{(k_n-k_l)(k_n-k_i)(k_l-k_i)(k_o^*-k_m^*)(k_o^*-k_j^*)(k_m^*-k_j^*)}
{D_2}\nonumber\\
&&\times\left|
\begin{array}{ccc}
\kappa_{ij} & \kappa_{im} & \kappa_{io}\\
\kappa_{lj} & \kappa_{lm} & \kappa_{lo} \\
\kappa_{nj} & \kappa_{nm} & \kappa_{no}
\end{array}
\right|
e^{\eta_i+\eta_j^*+\eta_l+\eta_m^*+\eta_n+\eta_o^*}\nonumber\\
&&+\frac{|k_1-k_2|^2|k_2-k_3|^2|k_3-k_1|^2|k_4-k_1|^2|k_2-k_4|^2|k_3-k_4|^2}
{\prod_{i=1}^{4}(k_i+k_i^*)|k_1+k_2^*|^2|k_1+k_3^*|^2
|k_1+k_4^*|^2|k_2+k_3^*|^2||k_2+k_4^*|^2|k_3+k_4^*|^2}\nonumber\\
&&\left|
\begin{array}{cccc}
\kappa_{11} & \kappa_{12} & \kappa_{13} & \kappa_{14}\\
\kappa_{21} & \kappa_{22} & \kappa_{23} & \kappa_{24}  \\
\kappa_{31} & \kappa_{32} & \kappa_{33} & \kappa_{34} \\
\kappa_{41} & \kappa_{42} & \kappa_{43} & \kappa_{44}
\end{array}
\right|
e^{(\eta_1+\eta_1^*+\eta_2+\eta_2^*+\eta_3+\eta_3^*+\eta_4+\eta_4^*)}.
\eear
In the above
\bear
D_2 &=& (k_j^*+k_i)(k_j^*+k_l)(k_j^*+k_n)(k_m^*+k_i)(k_m^*+k_l)\nonumber\\
&&(k_m^*+k_n)(k_o^*+k_i)(k_o^*+k_l)(k_o^*+k_n),
\eear
and 
\bear
\kappa_{il}=\frac{\mu\left(\al_i^{(1)}\al_l^{(1)*}+\al_i^{(2)}\al_l^{(2)*}
\right)}{(k_i+k_l^*)}, \;\; i,l=1,2,3,4.
\eear
\ees

\newpage
\section*{Figure Captions}
Fig. 1: Two distinct possibilities of the shape changing two soliton collision in the
integrable 2-CNLS system. The parameters are chosen as 
(a) $k_1=1+i$, $k_2=2-i$, $\al_1^{(1)}$ $=$ $\al_1^{(2)}=\al_2^{(2)}=1$,
 $\al_2^{(1)}=\frac{39+80i}{89}$; (b) 
$k_1=1+i$, $k_2=2-i$, $\al_1^{(1)}$ $=$ $0.02+0.1i$,
$\al_1^{(2)}=\al_2^{(1)}=\al_2^{(2)}=1$.

Fig. 2: Intensity profiles of the three modes of the two-soliton solution 
in a waveguide described by the 3-CNLS (Eq. (1) with $N=3$) showing  different 
dramatic 
scenarios of the shape changing collision for various choices of parameters.

Fig. 3 : Plot of the magnitude of phase shift as a function of the parameter 
$\al_1^{(1)}$, when it is real (for illustrative purpose), see Eqs. (29-31). The
other parameters are chosen as $k_1=1+i$, $k_2=2-i$, $\al_1^{(2)}=\al_2^{(2)}=1$
and $\al_2^{(1)}=\frac{39+80i}{89}$.

Fig. 4: A schematic three soliton collision process (for the choice $k_{1R}, k_{2R},
k_{3R} > 0$, $k_{1I}>k_{2I}>k_{3I}$). The effect of phase shifts are not included
in the figure.

Fig. 5: Intensity profiles $|q_1|^2$ and $|q_2|^2$ of the two modes of the 
three-soliton 
solution of the 2-CNLS equations, representing elastic collision, with the
parameters chosen as $k_1=1+i$, $k_2 = 1.5-0.5i$, $k_3 = 2-i$, 
$\al_1^{(1)}= \al_2^{(1)}= \al_3^{(1)}= \al_1^{(2)}=
\al_2^{(2)}=\al_3^{(2)}=1$.

Fig. 6: Intensity profiles  $|q_1|^2$ and $|q_2|^2$ of the two modes of the 
three-soliton 
solution of the 2-CNLS equations, representing the shape changing 
(intensity redistribution) collision process for the choice of the parameters, 
$k_1=1+i$, $k_2 = 1.5-0.5i$, $k_3 = 2-i$, 
$\al_1^{(1)}=\frac{39-80i}{89}$, $\al_2^{(1)}=\frac{39+80i}{89}$, 
$\al_3^{(1)}=0.3+0.2i$, $\al_1^{(2)}=0.39$, $\al_2^{(2)}=\al_3^{(2)}=1$.

Fig. 7: Shape restoring property of soliton 1 ($S_1$) during its collision with 
the other two solitons, soliton 2 ($S_2$) and soliton 3 ($S_3$), for the choice of
the parameters $k_1=1+i$, $k_2 = 1.5-0.5i$, $k_3 = 2-i$, 
$\al_1^{(1)}=\al_1^{(2)}=\al_2^{(2)}=1$, $\al_2^{(1)}=\frac{39+80i}{89}$, 
$\al_3^{(1)}=1.19$, $\al_3^{(2)}=\frac{39+80i}{89}$ .

Fig. 8: Typical 2-PCS forms for the Manakov system for $z=0$ see Eqs. (68),
with $k_1=1.0$ and $k_2=2.0$:
(a) Symmetric case ($t_{12}=0$), (b) Asymmetric case ($t_{12}=1$).

Fig. 9: Typical 3-PCS forms for the integrable 3-CNLS system for $z=0$ with
$k_1=1.0$, $k_2=0.5$ and $k_3=0.2$, see 
Eqs. (72):
(a) Symmetric case ($t_{12}= t_{13}=0$), (b) Asymmetric case ($t_{12}=1,\; 
t_{13}=2$).

Fig. 10: Intensity profiles showing the collision scenario of two 1-PCSs, with
equal velocities,
at (a) $z=-5$ and (b) $z=5$, given by special choice of parameters (as given in
text) in the 2-soliton solution of the Manakov system.

Fig. 11: Intensity profiles showing the collision scenario of two 1-PCSs,
moving with equal but opposite velocities, at (a) $z=-5$ and (b) $z=5$, given by special choice of parameters (as given in
text) in the 2-soliton solution of the Manakov system.

Fig. 12: Intensity profiles showing the collision scenario of 2-PCS with 1-PCS
at (a) $z=-4$ and (b) $z=4$ given by special choice of parameters (as given in
text) in the 3-soliton solution of Eq.(1) with $N=3$.

Fig. 13: Intensity profiles of a multisoliton complex comprising three solitons
spread up in two components propagating in photorefractive media: A special case
of the 
three soliton solution (10) of the integrable 2-CNLS system for the parameters
chosen as in the text, (a) at $z=-5$ and (b) at $z=5$.
\end{document}